\documentclass[aps,twocolumn,pre,longbibliography,floatfix,superscriptaddress]{revtex4-1}

\usepackage{amsmath,amsthm,amsfonts,amssymb,bm,graphicx,color,mathpazo,times, braket}
\usepackage[colorlinks={true}, citecolor={blue}, filecolor={blue}, linkcolor={blue}, urlcolor={blue}]{hyperref}
\usepackage[caption=false]{subfig}
\usepackage{soul}
\usepackage{amsmath}
\usepackage[mathlines]{lineno} 

\usepackage{xcolor}

\allowdisplaybreaks

\begin{document}
\title{Configuration-dependent precision in magnetometry and thermometry using multi-qubit quantum sensors}
\author{Asghar Ullah}
\email{aullah21@ku.edu.tr}
\affiliation{Department of Physics, Ko\c{c} University, 34450 Sar\i yer, Istanbul, T\"urkiye}
\author{\"Ozg\"ur E. M\"ustecapl\i o\u glu}
\email{omustecap@ku.edu.tr}
\affiliation{Department of Physics, Ko\c{c} University, 34450 Sar\i yer, Istanbul, T\"urkiye}
\affiliation{T\"UBİTAK Research Institute for Fundamental Sciences (TBAE), 41470 Gebze, Kocaeli, T\"urkiye}
\author{Matteo G. A. Paris}
\email{matteo.paris@fisica.unimi.it}
\affiliation{Dipartimento di Fisica, Università di Milano, I-20133 Milan, Italy}
\date{\today}

\begin{abstract}
We study the performance of quantum sensors composed of four qubits arranged in different geometries for magnetometry and thermometry. The qubits interact via the transverse-field Ising model with both ferromagnetic and antiferromagnetic couplings, maintained in thermal equilibrium with a heat bath under an external magnetic field. Using quantum Fisher information, we evaluate the metrological precision of these sensors. For ferromagnetic couplings, weakly connected graphs (e.g., the chain graph, $P_4$) perform optimally in estimating weak magnetic fields, whereas highly connected graphs (e.g., the complete graph, $K_4$) excel at strong fields. Conversely, $K_4$ achieves the highest sensitivity for temperature estimation in the weak-field regime. In the antiferromagnetic case, we uncover a fundamental trade-off dictated by spectral degeneracy: configurations with non-degenerate energy spectra—such as the pan-like graph (three qubits in a triangle with the fourth attached)—exhibit strong magnetic field sensitivity due to their pronounced response to perturbations. In contrast, symmetric structures like the square graph, featuring degenerate energy levels (particularly ground-state degeneracy), are better suited for precise thermometry. Notably, our four-qubit sensors achieve peak precision in the low-temperature, weak-field regime. Finally, we introduce a spectral sensitivity measure that quantifies energy spectrum deformations under small perturbations, providing a simple heuristic indicator of metrological sensitivity.
\end{abstract}

\maketitle
\section{Introduction}
Quantum metrology leverages quantum properties such as coherence~\cite{Wang_2018,PhysRevA.100.053825,PhysRevLett.107.083601} and entanglement~\cite{PhysRevA.94.012101,PhysRevLett.107.083601} to enhance measurement precision~\cite{PhysRevLett.96.010401,RevModPhys.89.035002,toth2014quantum,giovannetti2011advances,montenegro2024review}. Its applications range from biomedical imaging~\cite{Aslam2023,TAYLOR20161} to fundamental physics~\cite{DeMille2024}, pushing the boundaries of what can be reliably estimated. Temperature is a key physical parameter with broad relevance in physics, chemistry, and biology. Advanced thermal sensing techniques~\cite{Mehboudi_2019,Dedyulin_2022,PhysRevApplied.13.054057,PhysRevResearch.2.033389,PhysRevA.108.032220,Gottscholl2021}, including photonic~\cite{KLIMOV2018308,rossi2015entangled,Li,dong,bina2018continuous}, optomechanical~\cite{Purdy}, and cavity QED-based methods~\cite{Chowdhury_2019,PhysRevLett.125.120603,Galinskiy}, have been developed. Quantum approaches, such as critical~\cite{Aybar2022criticalquantum,PhysRevResearch.6.043094, PhysRevA.90.022111,PhysRevApplied.17.034073, Mok2021, PhysRevA.78.042105} and collisional thermometry~\cite{PhysRevLett.123.180602}, as well as those using cold atoms or spins~\cite{PhysRevLett.127.113602,PRXQuantum.3.040330,PhysRevLett.112.110403}, offer promising advancements. Similarly, quantum-enhanced magnetometry plays a crucial role in detecting weak magnetic fields with high precision~\cite{PhysRevX.5.031010, PhysRevA.99.062330, PhysRevLett.104.133601, PhysRevLett.120.260503, PhysRevLett.131.133602, PhysRevLett.129.120503, PhysRevA.99.062330,PhysRevA.109.032608, Li18, Albarelli_2017,PhysRevLett.134.010801}.

The structure and connectivity of quantum sensors are key to optimizing their performance~\cite{Abiuso_2024}, 
much like structured networks in biological and many-body quantum systems~\cite{Ishizaki_2012}. Nature exploits connectivity for enhanced functionality, and quantum networks use similar principles for sensing~\cite{PhysRevLett.120.080501,PhysRevResearch.3.033011} and information processing~\cite{PhysRevA.109.012424,Upadhyay_2024}. Extending these principles to quantum thermometry, it has been investigated recently \cite{PhysRevE.104.014136} 
how topology influences quantum thermometry precision by modeling the probe as a continuous-time quantum walker (CTQW), with the Hamiltonian corresponding to the graph Laplacian. Those results suggest that small algebraic connectivity is advantageous for precise temperature sensing.  In that reference, CTQW is used, which is directly related to the geometry of the spin network connectivity. However, the interplay between the underlying configuration geometry of the spins and the Hamiltonian of spin-spin interactions can be more complex in other models, such as the transverse-field Ising model. This distinction is critical for many applications, including quantum annealers~\cite{wolf,simon}. In our model system, we consider a transverse-field Ising model, and our optimum multi-qubit sensor configurations do not align with the categorizations found for CTQW models in Ref.~\cite{PhysRevE.104.014136}.

Building on these insights—but departing from the single-particle framework—we investigate the structural optimization of quantum sensors for both magnetometry and thermometry. Our quantum sensor is composed of four qubits arranged in six distinct connected graph-like configurations, where both the connectivity and arrangement of qubits play crucial yet subtle roles in sensing performance, as they shape the spectral structure of the network Hamiltonian and, in turn, influence the ultimate precision limits. The qubits interact via the transverse field Ising model, considering both ferromagnetic and antiferromagnetic couplings, and are coupled to a thermal bath under an external weak and uniform magnetic field, with the equilibrium state described by a Gibbs thermal state of the overall Hamiltonian. To quantify precision limits, we employ the quantum Fisher information (QFI), focusing on temperature and magnetic field estimation. We also introduce a spectral sensitivity measure that captures how the energy spectrum deforms under small perturbations, providing a simple heuristic for identifying configurations with enhanced quantum sensitivity.

In the ferromagnetic regime, we investigate how the structural configuration of the multi-qubit quantum sensor impacts the precision of parameter estimation. When estimating a weak magnetic field $h$, sensors with minimal connectivity, such as the linear-chain graph $P_4$, demonstrate optimal performance. Their low degree and simple energy spectra with degenerate ground states allow efficient spin flips at low fields, enhancing quantum fluctuations and yielding a large QFI. The star graph $S_3$ similarly performs well due to comparable spectral properties. In contrast, highly connected structures like the complete graph $K_4$ are stiffer, resisting spin flips and thus requiring stronger magnetic fields to achieve significant sensitivity. As a result, graphs with higher connectivity outperform at stronger magnetic fields, although the maximum achievable precision slightly decreases. For temperature estimation, the situation is somewhat different. The optimal performance depends on both the structure and the type of ferromagnetic or antiferromagnetic coupling. The complete graph $K_4$ shows the highest thermal sensitivity, benefiting from its highest connectivity, which balances energy level structure and thermal susceptibility. The connection between the thermal QFI and the heat capacity emphasizes that structures with a spectrum favoring low-energy excitations yield better performance at low temperatures, but stiffer graphs maintain better sensitivity under stronger external fields.

In the antiferromagnetic case, our findings indicate that the optimal configuration for magnetic field estimation is a pan graph (a triangle with a pendant vertex). The precision of magnetic sensing is closely linked to the eigenvalue spectrum of the system's Hamiltonian. Intuitively, unless special entangled states are used to exploit degeneracies, non-degenerate energy levels generally enhance magnetic sensitivity, whereas degenerate levels limit the system’s ability to distinguish between energy levels under magnetic perturbations. The presence of degeneracies, particularly ground-state degeneracy, in the spectrum reduces the contrast in the system's response to changes in the magnetic field, thereby diminishing its ability to extract information from magnetic perturbations. Among the considered configurations, the pan graph (and $K_4$) exhibits fewer symmetries in the energy spectrum compared to the others, resulting in a highly non-degenerate spectrum, with no ground state degeneracy, which is particularly advantageous for magnetic sensing. Notably, among the 4-qubit sensor configurations, the pan graph possesses a non-degenerate spectrum at finite magnetic field, making it the most favorable choice for precision magnetometry. In contrast, the $C_4$ graph configuration is the most effective for temperature estimation, as its highly symmetric energy spectrum exhibits degenerate ground states in the weak-field regime, and consequently offers the highest sensitivity to temperature variations. This result aligns with existing literature, where it is well established that degeneracy enhances thermal sensitivity~\cite{campbell2018precision,PhysRevLett.114.220405,PhysRevA.97.063619,topological_thermometry,PhysRevA.111.052216}. In addition, among the six distinct connected graphs for a four-qubit sensor, the square configuration demonstrates the sharpest Boltzmann factor variation with temperature, enabling rapid state population changes and superior temperature sensitivity compared to more gradual-response alternatives. Notably, the pan and $K_4$ graphs are the only configurations that do not have ground state degeneracy, and the energy levels are more shifted under magnetic perturbation, making them good magnetometers. In contrast, the other configurations, with ground state degeneracy, are more effective for temperature sensing, making them better suited as thermometers.

Overall, the sensor's energy spectrum, connectivity, and degree of symmetry are critical factors in optimizing quantum sensing: sparse graphs offer superior sensitivity for weak-field magnetometry, while more connected graphs provide robustness under strong fields, enhancing both high-field magnetometry and thermometric precision.

The remainder of the paper is structured as follows. In Sec.~\ref{PET}, we introduce the fundamental concepts of quantum parameter estimation theory. Sec.~\ref{model} provides an overview of the model used in this study. The results for quantum magnetometry and thermometry are discussed in Sec.~\ref{results}. In Sec.~\ref{deform} we introduce a spectral deformation measure that offers qualitative insight into which sensor configurations may exhibit stronger magnetic field sensitivity. Section~\ref{conc} concludes with a summary of our findings. We discuss the role of the external magnetic field on the energy spectrum of each configuration in Appendix~\ref{appA}, while the effects of temperature on the precision of magnetic field estimation, and vice versa, are presented in Appendices~\ref{effectT} and~\ref{effecth}, respectively. 
\section{Parameter estimation theory}\label{PET}
In this section, we provide a concise overview of quantum estimation theory, focusing on single-parameter estimation. We introduce key concepts and tools that will be used throughout this study.
In quantum metrology, the maximum achievable precision about an unknown parameter $\theta$ is quantified by quantum Fisher information (QFI). For any parameterized quantum state $\rho_\theta$ , the ultimate precision limit in estimating the parameter $\theta$ is determined by the Cramér-Rao bound such that the variance of any estimator $\hat{\theta}$ is lower bounded by the reciprocal of the QFI. The mathematical expression is given by~\cite{helstrom1969quantum,PhysRevLett.72.3439,paris2009quantum,sidhu}
\begin{equation}
    \mathrm{Var}(\theta)\ge\frac{1}{mF_Q(\theta)},
\end{equation}
where $\mathrm{Var}(\theta)$ denotes the variance, $m$ represents the number of measurements repeated, and $F_Q(\theta)$ is the QFI, which depends on the quantum state $\rho_\theta$ and its parameterization process. The QFI is maximized over all possible POVMS~\cite{PhysRevLett.72.3439} to obtain the ultimate bound on the precision of parameter $\theta$. For a mixed state, the QFI is defined as
\begin{equation}
    F_Q(\theta)=\text{Tr}[\rho_\theta L_\theta^2],
\end{equation}
where $L_\theta$ is the symmetric logarithmic derivative (SLD), a Hermitian operator which is explicitly defined by
\begin{equation}
    \frac{\partial\rho_\theta}{\partial\theta}=\frac{L_\theta\rho_\theta+\rho_\theta L_\theta}{2}.
\end{equation}
If we express $L_\theta$ in the eigenbasis of $\rho_\theta$, the QFI can be written as follows:
\begin{equation}\label{qfi}
F_Q(\theta) = 2 \sum_{n,m} \frac{|\langle \psi_n | \partial_\theta \rho_\theta | \psi_m \rangle|^2}{\lambda_n + \lambda_m},
\end{equation}
where $\lambda_n$ and $\lambda_m$ are the eigenvalues and $|\psi_n\rangle$ and $|\psi_m\rangle$ represent the eigenvectors of the density matrix $\rho_\theta$ and $\partial_\theta=\partial/\partial_\theta$ is the partial derivative with respect to $\theta$.

We remark that the SLD is defined based on the density matrix $\rho_\theta$ and both the eigenvalues $\lambda_n$ and eigenvectors $|\psi_n\rangle$ may depend on the parameter $\theta$. Therefore, we can separate the contribution of QFI into two parts by using the Leibniz rule:
\begin{equation}\label{rhoD}
\begin{aligned}
\partial_\theta\rho_\theta &= \sum_n \big( (\partial_\theta \lambda_n) |\psi_n\rangle \langle \psi_n| \\
&\quad + \lambda_n |\partial_\theta \psi_n\rangle \langle \psi_n| + \rho_\theta |\psi_n\rangle \langle \partial_\theta \psi_n| \big),
\end{aligned}
\end{equation}
The symbol $|\partial_{\theta} \psi_n\rangle$ denotes the ket, such as
\begin{equation}
    |\partial_{\theta} \psi_n\rangle = \sum_k \partial_{\theta} \psi_{nk} |k\rangle,
\end{equation}
where $\psi_{nk}$ are obtained by expanding $|\psi_n\rangle$ in an arbitrary basis $\{|k\rangle\}$ independent of $\theta$.  
Since $\langle \psi_n | \psi_m \rangle = \delta_{nm}$, we have
\begin{equation}
    \partial_{\theta} \langle \psi_n | \psi_m \rangle \equiv \langle \partial_{\theta} \psi_n | \psi_m \rangle + \langle \psi_n | \partial_{\theta} \psi_m \rangle = 0
\end{equation}
and therefore
\begin{equation}
    \text{Re} \langle \partial_{\theta} \psi_n | \psi_m \rangle = 0, \quad \langle \partial_{\theta} \psi_n | \psi_m \rangle = -\langle \psi_n | \partial_{\theta} \psi_m \rangle = 0.
\end{equation}
Using Eq.~\eqref{rhoD} and the above identities, we have
\begin{equation}
\begin{aligned}
L_{\theta} &= \sum_p \frac{\partial_{\theta} \lambda_p}{\lambda_p} |\psi_p\rangle \langle \psi_p| \\
&+2 \sum_{n\neq m} \frac{\lambda_n - \lambda_m}{\lambda_n + \lambda_m} \langle \psi_m | \partial_{\theta} \psi_n \rangle |\psi_m\rangle \langle \psi_n| 
\end{aligned}
\end{equation}
and in turn, the QFI takes the form
\begin{equation}\label{qfiFull}
F_Q(\theta) = \sum_p \frac{(\partial_{\theta} \lambda_p)^2}{\lambda_p} + 2 \sum_{n\neq m}\frac{(\lambda_n - \lambda_m)^2}{\lambda_n + \lambda_m} |\langle \psi_m | \partial_{\theta} \psi_n \rangle|^2.
\end{equation}
The first term in Eq.~\eqref{qfiFull} represents the classical contribution to the QFI, while the second term captures the purely quantum contribution. In this work, we focus on single-parameter estimation, exploiting the field as an external tuning to possibly enhance thermometry, and temperature as a source of noise in the estimation of the magnetic field. Joint estimation of the two quantities may be investigated further, but it is beyond the scope of the present work. 

\begin{figure}[t!]
    \centering
    \includegraphics[scale=0.8]{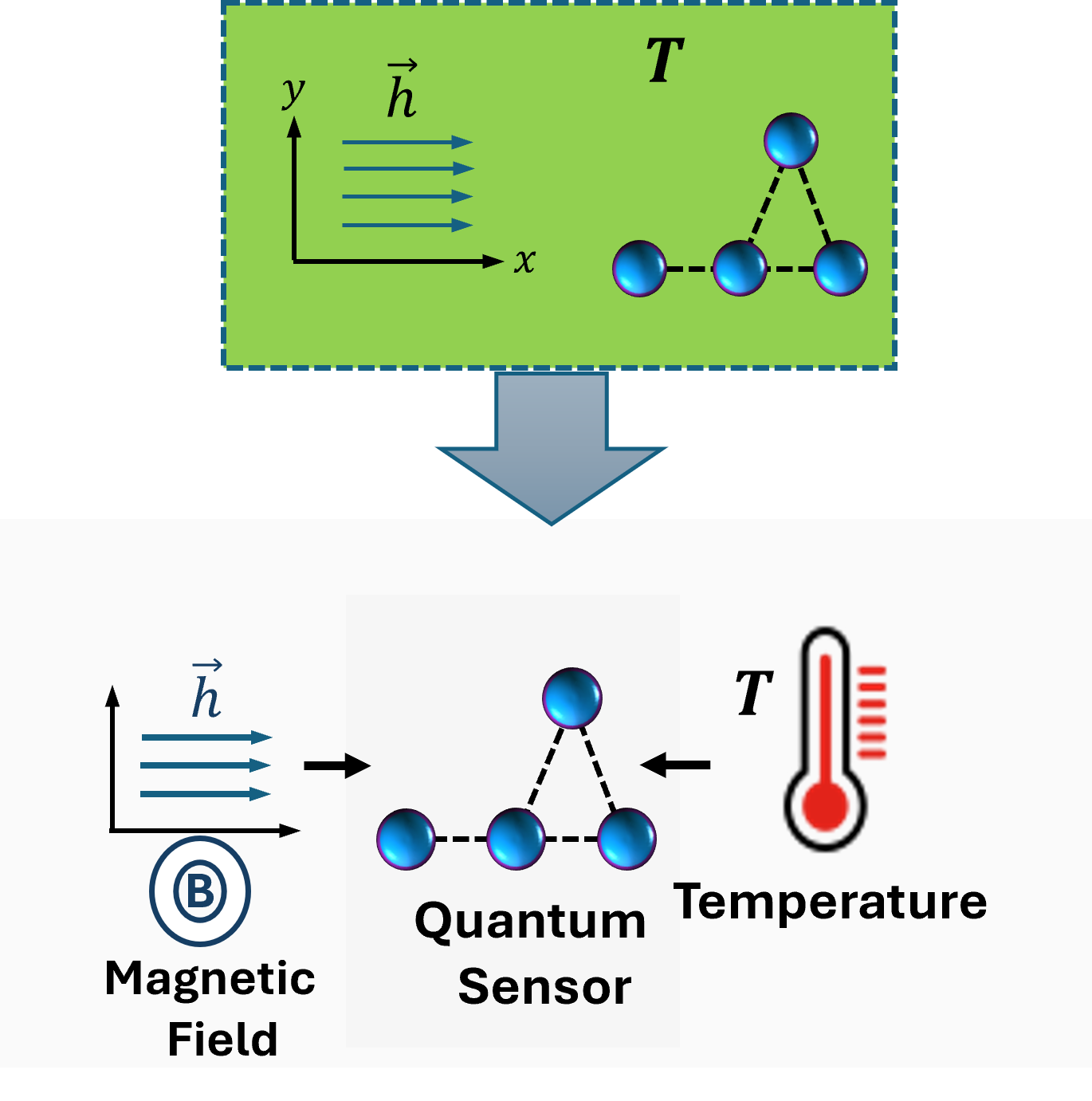}
    \caption{Model representation of our quantum sensing protocol. The quantum sensor, modeled as a network of four coupled spins, simultaneously exposed to a thermal bath at temperature \( T \) and an external weak magnetic field with magnitude \( \Vec{h} \), directed along $x$-axis. The sensor functions as a probe to estimate unknown parameters, including $T$ and $h$. The different configurations of the quantum sensor are given in Fig.~\ref{fig2}.
}\label{fig1}
\end{figure}
\section{The Model}\label{model}
In this work, we consider a finite-size quantum sensor consisting of four 
 coupled spin-\(1/2\) particles, such as qubits, coupled via an Ising spin chain in the presence of an external magnetic field. We assume that the sensor is in thermal equilibrium with a thermal bath at temperature \(T\). The sensor is used to measure both the magnetic field and temperature (see Fig.~\ref{fig1}). 

\begin{figure}[t!]
    \centering
    \includegraphics[scale=0.64]{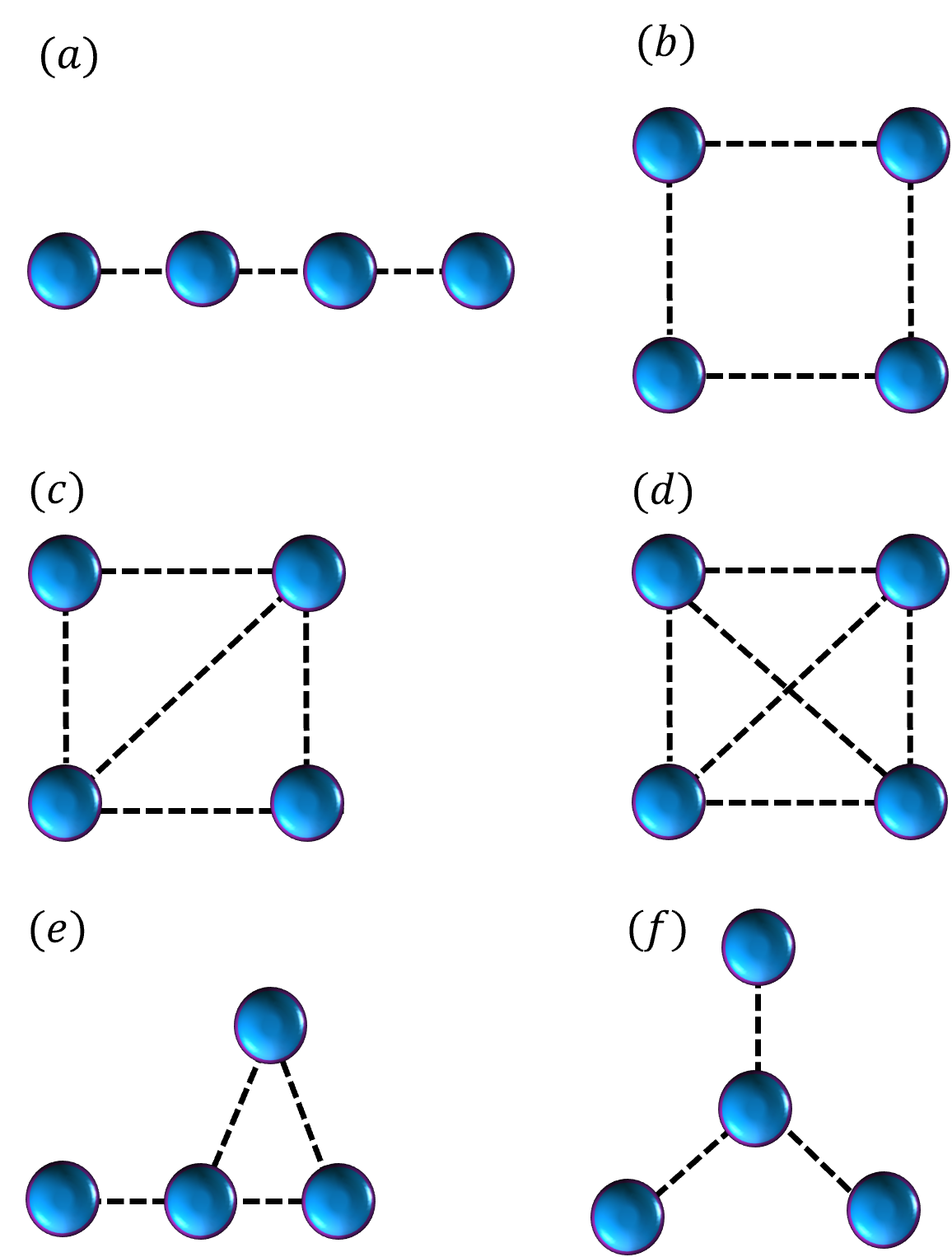}
    \caption{Graphical representation of the structural arrangement of four spins in six different configurations: (a) Chain ($P_4$), (b) $4$-cycle square ($C_4$), (c) $4$-cycle with a diagonal ($Sd_4$), (d) Complete ($K_4$), (e) Triangle with a pendant vortex (pan graph), (f) Star-tree graph ($S_3$). In each configuration, the spins are coupled to a thermal bath at temperature \( T \) and subjected to an external magnetic field of strength \( h \). The vertices, represented by solid purple spheres, denote the spin qubits, while the dashed lines indicate the interactions between them.}
    \label{fig2}
\end{figure}

The four spins can be arranged in six different configurations, assuming only connected graphs. These include a linear chain, a complete cycle square, a cycle square with one and two diagonal interactions, a triangle with a pendant vortex, and a star-tree graph, as shown in Figs.~\ref{fig2}(a)-(f), respectively.

\begin{table}[b!]
\centering
\begin{tabular}{c lcccc}
\hline\hline
Index & Graph & No. of Edges & Total degree & Ground state degeneracy \\
\hline
(a) & \(P_4\) & 3 & 6  & 2 \\
(b) & \(C_4\) & 4 & 8  & 2 \\
(c) & $Sd_4$ & 5 & 10  & 2 \\
(d) & \(K_4\) & 6 & 12  & 6 \\
(e) & Pan & 4 & 8  & 6 \\
(f) & \(S_3\) & 3 & 6 & 2 \\
\hline\hline
\end{tabular}
\caption{Properties of the six graph configurations. The "Degree" refers to the sum of the degrees of all nodes, and "Degeneracy" refers to the number of distinct ground state degeneracies in the absence of an external magnetic field ($h=0$) and for the antiferromagnetic coupling $J = -1$}.
\label{fig:properties}
\end{table}
The Hamiltonian of the quantum sensor composed of $N=4$ qubits governed by the transverse field Ising model is given by (we set \(\hbar = k_B = 1\))~\cite{Fischer_Hertz_1991,landau2013statistical}:  
\begin{equation}
\hat{H}_S = -\sum_{i>j=1}^{4} J_{ij} \hat{\sigma}_i^z \hat{\sigma}_j^z - h_x \sum_{i=1}^{4} \hat{\sigma}_i^x,
\end{equation}
where \( J_{ij} \) represents the coupling strength between qubits, \(\hat{\sigma}_i^z\) and \(\hat{\sigma}_j^z\); \(\hat{\sigma}^\alpha\) (\(\alpha = x, y, z\)) are the Pauli operators; and \( h_x \) denotes the strength of the external magnetic field. For simplicity, we assume homogeneous coupling such that \(J_{ij} = J\) and homogeneous magnetic field strength \(h_x = h\) throughout the paper.

The sign of the coupling constant \(J\) determines whether the system exhibits ferromagnetic or antiferromagnetic behavior:
\begin{itemize}
    \item For ferromagnetic coupling (\(J > 0\)), the qubits tend to align in the same direction, promoting long-range ferromagnetic order.
    \item For antiferromagnetic coupling (\(J < 0\)), the qubits prefer to align in opposite directions, favoring a more disordered or anti-aligned state.
\end{itemize}
In this study, we explore both ferromagnetic (\(J = 1\)) and antiferromagnetic (\(J = -1\)) configurations, considering their effects on quantum sensor performance.

The goal is to use this quantum sensor as a probe to measure the temperature \( T \) of the thermal bath and the external magnetic field \( h \) (see Fig.~\ref{fig1}). Notably, our quantum sensing scheme does not rely on any external drive or strong nonlinear coupling to the environment. Instead, we consider a standard thermodynamic scenario, where a system in contact with a thermal bath at temperature \( T \) undergoes thermalization and eventually reaches equilibrium at the same temperature~\cite{PhysRevLett.134.010801,PhysRevLett.114.220405}. While many quantum probing schemes focus on extracting parameter information from non-equilibrium dynamics~\cite{PhysRevE.110.024132, Sekatski2022optimal,PhysRevA.98.050101, Sekatski2022optimal, PhysRevA.99.062114, PhysRevA.108.062421, PhysRevResearch.2.033498, Razavian2019, PhysRevLett.125.080402, Ravell, Paris_2016}, our work considers quantum sensors at thermal equilibrium—a setting that, although potentially less optimal in speed or coherence usage, aligns more closely with realistic experimental conditions. 
\begin{figure*}[t!]
    \centering
    \includegraphics[scale=0.44]{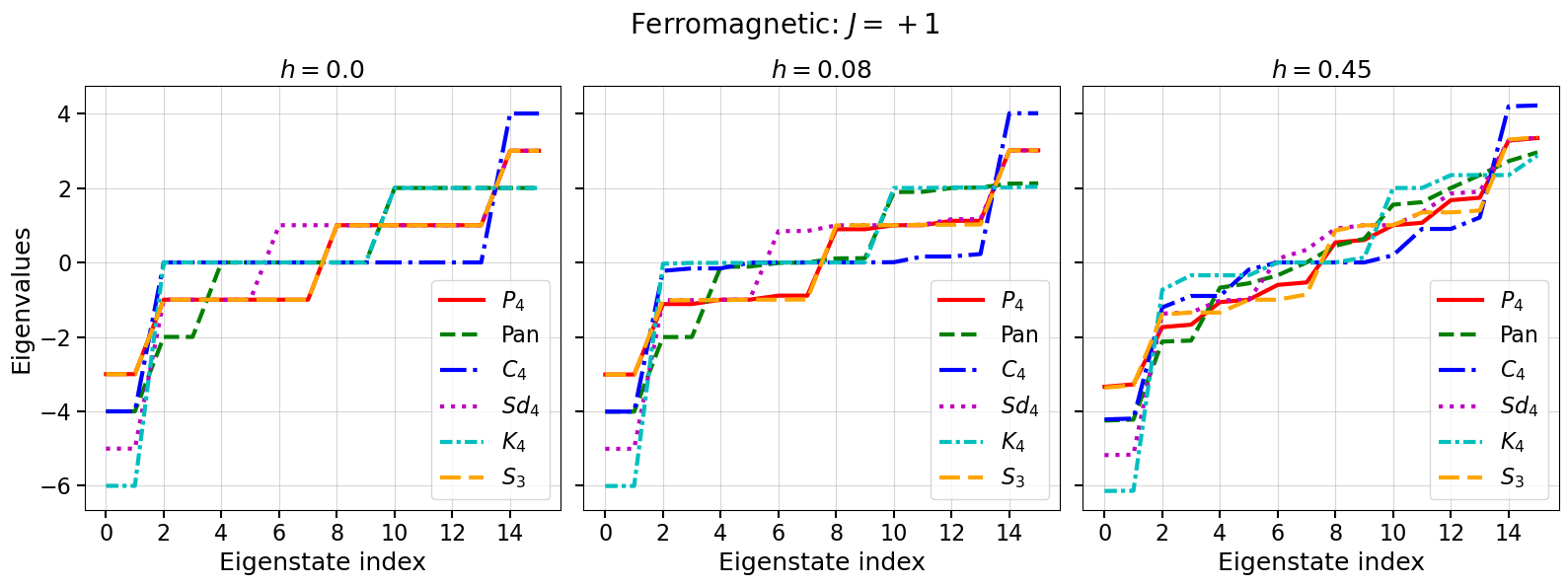}\\
    \includegraphics[scale=0.44]{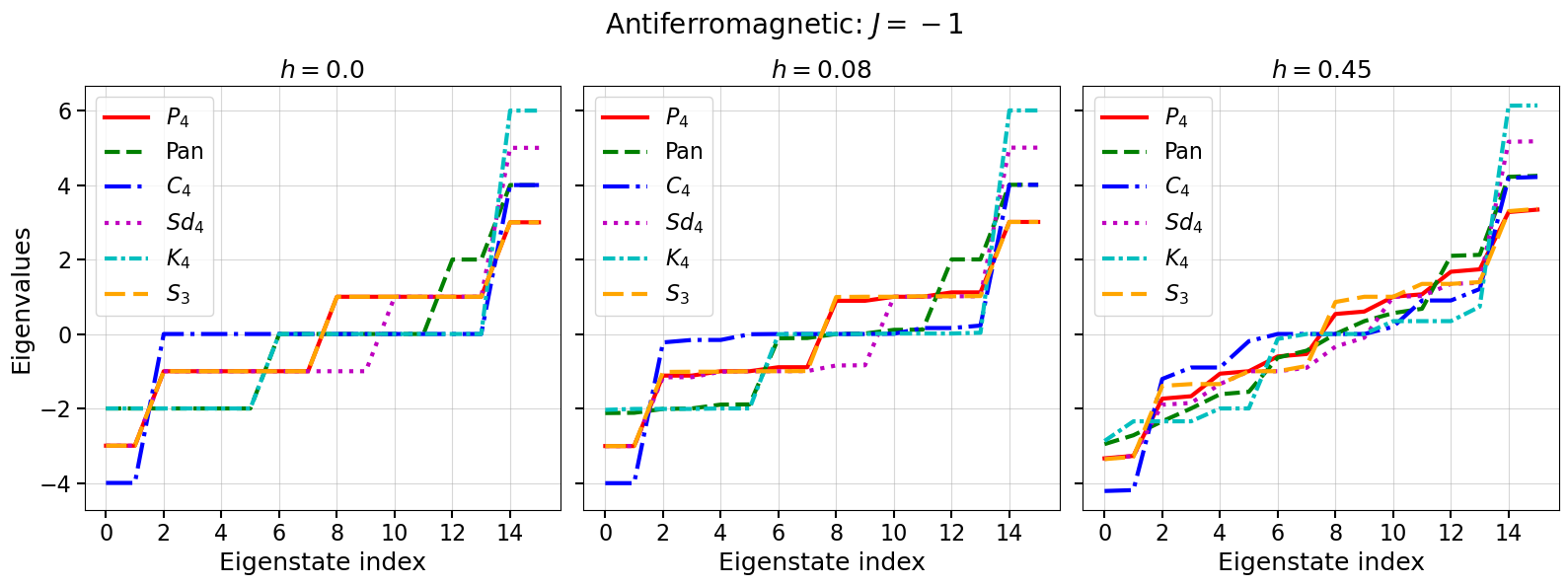}
    \caption{Energy eigenvalue spectra for six different configurations of the quantum sensor in the presence of a weak external magnetic field (\(h = 0.08\)). The top panel corresponds to ferromagnetic coupling (\(J = 1\)) and the bottom panel to antiferromagnetic coupling (\(J = -1\)). The legends refer to six different configurations of the $4-$qubit quantum sensor, which are given in Fig.~\ref{fig2}.}
    \label{fig:Eb}
\end{figure*}

We assume that the system is weakly coupled to a low-temperature bath and subjected to a weak external magnetic field. Under these conditions, the sensor reaches thermal equilibrium with the bath while experiencing the field \( h \). Consequently, the system's state can be well described by the Gibbs thermal state:
\begin{figure}[t!]
    \centering
    \subfloat[]{
        \includegraphics[scale=0.55]{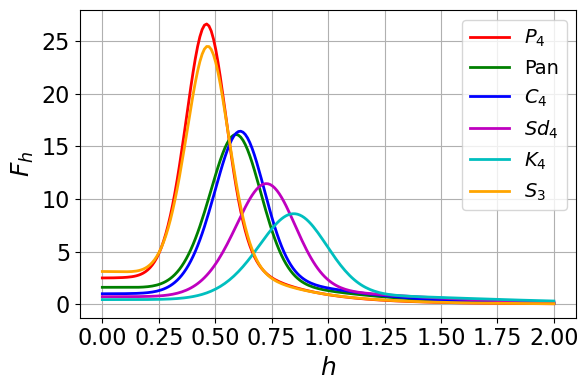}} \\ 
        \subfloat[]{
        \includegraphics[scale=0.55]{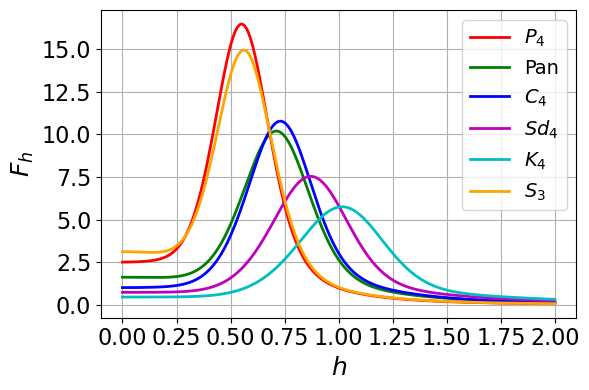}} 
    \caption{The QFI ($F_h$) for magnetic field ($h$), is plotted as a function of the magnetic field strength $h$ for different quantum sensor configurations at temperatures \( T = 0.04 \) (a) and \( T = 0.08 \) (b). The different colors, such as red, green, blue, magenta, cyan, and orange, represent the configurations \( P_4 \), \( C_4 \), Pan, \( Sd_4 \), \( K_4 \), and \( S_3 \), respectively. All plots correspond to the ferromagnetic coupling (\( J = 1 \)).
    }
    \label{fig3}
\end{figure}
\begin{equation}
    \rho_G(T,h) = \frac{\text{exp}{(-\beta \hat{H}_S)}}{\mathcal{Z}(T,h)},
\end{equation}
where \( \beta = 1/T \) (we set \( k_B = 1 \)) is the inverse temperature and 
\begin{equation}
    \mathcal{Z}(T,h) = \text{Tr}\big[\text{exp}(-\beta \hat{H}_S)\big] 
\end{equation}
is the partition.

In the current setting, the QFI is used to analyze the sensor's ability to individually estimate both the magnetic field \( h \) and the temperature \( T \). We numerically compute the QFI using the eigenvalues and eigenvectors of \( \rho_G(T,h) \) to quantify the precision of parameter estimation for both \( h \) and \( T \)~\cite{PhysRevResearch.4.043057}.  
\section{Results }\label{results}
In this section, we present the results of our quantum sensing scheme for the estimation of the magnetic field $h$ and thermal bath temperature $T$. We perform exact diagonalization of the Hamiltonian $\hat{H}(h)$ to obtain its eigenvalues $E_n$ and eigenvectors $|\psi_n\rangle$ and then construct the thermal state where the populations are given by $p_n=e^{-\beta E_n}/\mathcal{Z}$. The derivative of the eigenvectors with respect to a parameter $\theta$ are evaluated by taking the derivative the characteristic equation $\hat{H} |\psi_n\rangle = E_n |\psi_n\rangle$ and rearranging terms, arriving at
\begin{equation} |\partial_\theta \psi_n\rangle = \sum_{m\neq n} \frac{\langle \psi_m | \partial_\theta \hat{H} | \psi_n \rangle}{E_n - E_m} |\psi_m\rangle, 
\end{equation} 
where $\partial_\theta\hat{H}$ is known from the model (i.e., $-\sum_i\hat{\sigma}^x$ in the transverse field Ising model).
For completeness, we note that in all numerical calculations, the QFI is evaluated at strictly finite temperature $T>0$, ensuring that the Gibbs state $\rho_G(T,h)$ is full rank and that all eigenvalues $\lambda_n$ are strictly positive. In addition, we do not evaluate the QFI exactly at $h=0$, where the Hamiltonian spectrum contains exact degeneracies; instead, we use a small symmetry-breaking field $h\to 0^{+}$ to lift these degeneracies. In implementing Eq.~\eqref{qfiFull}, we follow the standard mixed-state QFI definition, according to which terms with $\lambda_n+\lambda_m=0$ do not contribute and are omitted. This ensures that the numerical evaluation of the QFI remains well-defined and stable throughout. Consequently, in our finite-size and finite temperature setting, the QFI shows enhanced but never singular behavior as a function of $h$, in contrast to the genuine critical divergences expected only in the thermodynamic limit~\cite{PhysRevA.78.042105}.
\subsection{Ferromagnetic sensors of magnetic field}
We first assume that the temperature of the thermal bath is known and use Eq.~\eqref{qfiFull} to calculate the QFI for the estimation of the magnetic field $h$. We set the interaction between the qubits to $J=1$. Figure~\ref{fig3} presents the QFI as a function of the magnetic field $h$ for the different sensor configurations shown in Fig.~\ref{fig2}. Each curve in Fig.~\ref{fig3} corresponds to a specific structural arrangement of the four-qubit sensor, allowing for a comparative analysis of how different configurations influence the estimation precision of $h$.

In the linear-chain configuration ($P_4$), the four qubits are arranged in a 1D array with nearest-neighbor interactions. This structure exhibits the largest response to the weak magnetic field, with a gradual increase in QFI as $h$ increases, as shown by the red curve in Fig.~\ref{fig3}(a). The QFI of the star graph ($S_3$) is slightly lower than that of $P_4$ (orange curve) because both graphs have the same energy spectrum, in particular ground states in the weak field regime, as shown in Fig.~\ref{fig:Eb} (top panel). In both cases, the QFI rises sharply at a certain value of the magnetic field $h$, reaching the maximum value $F_{\text{max}}$.  
Both graphs are less symmetric, with only three edges, and possess a simple energy spectrum with 2-fold ground-state degeneracy (see Fig.~\ref{fig:Eb} for $h=0$ and Table~\ref{fig:properties} for graphs' properties), making it easier to flip spins at small $h$ because the graphs are more fragile. At small $h$, the system is dominated by the Ising interaction. Due to the minimal connectivity of $P_4$ and $S_3$, local spin flips cost relatively little energy, resulting in many accessible low-energy states. Even a weak transverse field efficiently mixes these states, creating coherent superpositions (quantum fluctuations) and leading to a large QFI.  
The magnetic sensitivity is highest at low fields when the temperature is low; however, it decreases, and the maximum slightly shifts toward higher magnetic fields as the temperature increases, as quantum fluctuations become washed out. The increase in temperature helps to melt the fragility of the graphs, making the spin flip easier. This means that higher temperatures degrade the metrological performance in estimation of $h$ (see Fig.~\ref{fig:effectT} in Appendix~\ref{effectT} for more details).

On the other hand, $C_4$ configuration (Fig.~\ref{fig2}(b)) where the four qubits form a square lattice with interactions only between nearest-neighbor pairs along the edges, excluding diagonal couplings. While Fig.~\ref{fig2}(e) represents the triangle with pendant vertex (pan graph) configuration, where three qubits form a triangular structure with the fourth qubit attached via a single interaction. Interestingly, these two configurations give nearly the same magnetic sensing precision shown by the blue and green curves in Fig.~\ref{fig3}(b), respectively. This is possible because both graphs are more connected than the chain or star graphs with the same edges, degree, and ground-state degeneracy. As a result, the local spin flips become costly in energy. The maximum of QFI also slightly shifts towards the higher magnetic field.

Adding one additional diagonal interaction to the square configuration creates a 4-cycle square with a diagonal ($Sd_4$). As shown by the magenta curve in Fig.~\ref{fig3}(b), the QFI further decreases due to the extra connection, shifting the maximum towards a higher magnetic field. If the configuration has even more connections, such as the complete graph ($K_4$) with all possible diagonal interactions (see Fig.~\ref{fig2}(d)), the Ising part couples the spins strongly, even for large \( h \). This high connectivity of the graph with the largest degree generates nontrivial correlations that persist against the magnetic field and make the graph stiffer. Therefore, this graph is better for enhancing the magnetic precision at higher values of $h$. Although increasing the temperature reduces the sensitivity in all cases, however, the overall collective behavior remains similar (as shown in Fig.~\ref{fig:effectT} of Appendix~\ref{effectT}).

\begin{figure}[t!]
    \centering
         \subfloat[]{
        \includegraphics[scale=0.55]{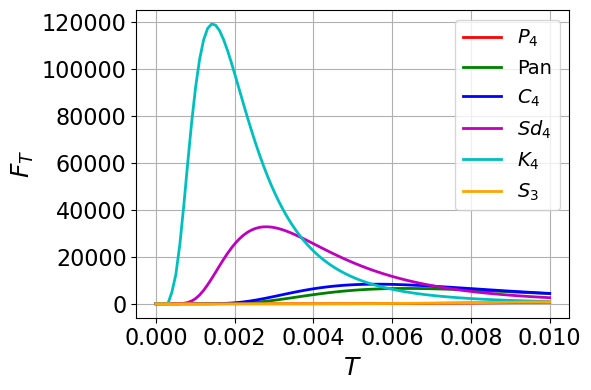}} \\
         \subfloat[]{
        \includegraphics[scale=0.55]{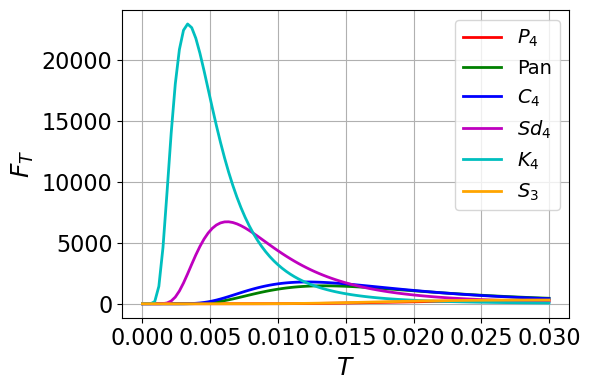}} 
    \caption{The QFI ($F_T$) temperature estimation is plotted as a function of temperature $T$ for two values of magnetic fields $h = 0.45$  (a) and $h = 0.55$ (b).  The different colors, such as red, green, blue, magenta, cyan, and orange, represent the configurations \( P_4 \), \( C_4 \), Pan, \( Sd_4 \), \( K_4 \), and \( S_3 \), respectively. All plots correspond to the ferromagnetic coupling (\( J = 1 \)).
    }
    \label{fig4}
\end{figure}

These results show that when the coupling between spins is ferromagnetic ($J=1$), the connectivity and symmetry of the sensor configuration play an important role in determining performance (see Table.~\ref{fig:properties} for more details). We observe that a sensor with minimal connectivity, such as a chain, is optimal for weak magnetic field estimation (low $h$), whereas a complete graph with more connections is better for high-field sensitivity, although the maximum precision is somewhat reduced. The QFI $F_h$ for the estimation of the magnetic field $h$ is linked with the magnetic susceptibility $\chi_M$ ($F_h\sim\chi_M$), which means that the larger the $\chi_M$, the greater will be the QFI yielding the maximum precision for magnetic field estimation.

\subsection{Ferromagnetic thermometers}
In the previous section, we employed the quantum sensor as a magnetometer and compared the performance of different sensor configurations in estimating the magnetic field with high precision. In this section, we assume that the magnetic field is known and instead, we utilize the sensor as a thermometer to measure the temperature of the thermal bath, and we analyze the effectiveness of various structural configurations in temperature estimation.
As we assume that the sensor is at thermal equilibrium with the bath of temperature $T$ in the presence of the external magnetic field $h$, we can also use thermal QFI $F_{th}$ to calculate the QFI for a thermal state $\rho_{th}$ which can be linked to the heat capacity, and hence it turns out to proportional the variance of the sensor Hamiltonian that is given by~\cite{Mehboudi_2019}
\begin{equation}
    F_{th}=\frac{\Delta\hat{H}_p^2}{T^4}=\frac{\langle\hat{H}_p^2\rangle-\langle\hat{H}_p\rangle^2}{T^4},
\end{equation}
where $\langle\hat{H_p}\rangle=\text{Tr}[\rho_{th}\hat{H}_p]$ represents the expectation of the value of the probe Hamiltonian $\hat{H_p}$ for a thermal state $\rho_{th}$.
It is important to highlight that the Cramér-Rao bound is saturated at any temperature when the probe is measured in the energy eigenbasis~\cite{PhysRevA.82.011611}, where the thermal state $\rho_{th}$ remains diagonal. Consequently, in equilibrium thermometry, enhancing precision is only possible by increasing the thermal QFI through careful engineering of the probe.

The QFI as a function of temperature $T$ for various sensor configurations is illustrated in Fig.~\ref{fig4} for two values of temperature. We investigate the influence of the magnetic field $h$ on the precision of temperature estimation by considering two different values of $h$, focusing on the ferromagnetic case where $J=1$. Our analysis indicates that the $K_4$ graph exhibits the highest sensitivity for temperature estimation, followed by the $Sd_4$ and $C_4$ structures. This behavior can be attributed to their respective energy spectra, as illustrated in Fig.~\ref{fig:Eb}, where the $K_4$ configuration has the lowest ground-state energy, followed by $Sd_4$ and then $C_4$. Notably, their precision in temperature estimation aligns with this ordering, suggesting a correlation between the ground-state energy structure and thermometric performance. We note that the temperature precision is maximal at very low temperatures and decreases rapidly as the temperature increases (see Fig.~\ref{fig4}(a)).

Furthermore, we find that stronger magnetic fields have a detrimental effect on temperature estimation precision (see Appendix~\ref{effecth}). Notably, when $h=0.55$, the highest sensitivity is still exhibited by $K_4$, as shown in Fig.~\ref{fig4}(b). It is interesting to observe that the $K_4$ graph, shown in Fig.~\ref{fig2}(d), features the maximum number of interactions among the qubits. For instance, the graphs with a greater number of edges and degree are stiffer (see properties of $Sd_4$ and $K_4$ in Table.~\ref{fig:properties}) and yield greater thermometric precision in the presence of a strong external magnetic field.

\subsection{Antiferromagnetic sensors of magnetic field}
\begin{figure}[t]
    \centering         
    \subfloat[]{
        \includegraphics[scale=0.55]{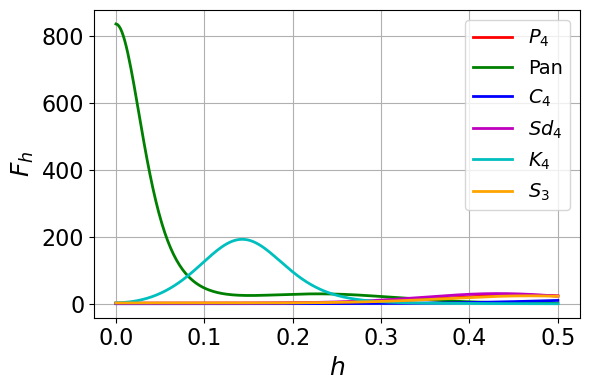}} \\
        \subfloat[]{
        \includegraphics[scale=0.55]{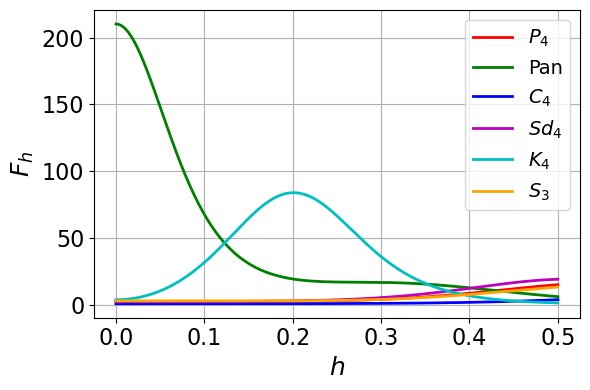}} 
    \caption{The QFI ($F_h$) for magnetic field estimation is plotted as a function of the magnetic field strength $h$ for different quantum sensor configurations at temperatures \( T = 0.04 \) (a) and \( T = 0.08 \) (b). The different colors, such as red, green, blue, magenta, cyan, and orange, represent the configurations \( P_4 \), \( C_4 \), Pan, \( Sd_4 \), \( K_4 \), and \( S_3 \), respectively. All plots correspond to the antiferromagnetic coupling (\( J = -1 \))}
    \label{fig5}
\end{figure}

We now consider the case of antiferromagnetic couplings such as $J=-1$ and show the results in Fig.~\ref{fig5} for the estimation of the magnetic field for two values of $T$.

It is immediately evident that the QFI yields entirely different results when the sign of $J$ is reversed. In this case, the pan graph exhibits the highest QFI for magnetic field estimation (green curve in Fig.~\ref{fig5}(a)), while the complete graph ($K_4$) ranks second (cyan curve in Fig.~\ref{fig5}(a)). 
The high precision achieved by the pan configuration lies in the weak magnetic field regime, whereas $K_4$ shows better sensitivity at higher magnetic fields due to its greater connectivity when $T=0.04$ in both cases. 
However, when the temperature is increased to $T=0.08$, the precision significantly decreases (see Appendix~\ref{effectT} for the effect of $T$ on the estimation precision of $h$). 
Of particular importance is the fact that while the other graphs also exhibit nonzero QFI, their values remain much smaller compared to the pan and $K_4$ sensors.

The variation in magnetic field precision across different structures can be understood by analyzing the eigenvalue spectra of each network. To illustrate this, we compare the spectra of two graphs: the $C_4$ structure, which exhibits the lowest QFI, and the pan graph, which achieves the highest QFI for magnetic field sensing. Fig.~\ref{fig:Eb} presents the eigenvalue spectra for these two graphs for different values of $h$ when $J=-1$ (see bottom panel).

In Fig.~\ref{fig:Eb} (bottom panel), we present the eigenvalue spectrum for all six graphs when \( h = 0, 0.08, 0.45 \) and $J=-1$. For $h=0$, the $C_4$ has only the two lowest degenerate energy levels, with most of the zero-energy levels that are also degenerate. Upon applying the external magnetic field (such as $h=0.45$), these energy levels shift, as shown in Fig.~\ref{fig:Eb} (bottom panel). The eigenvalue spectrum of \( C_4 \) reveals four eigenvalues that still remain degenerate at zero energy, which may not have a significant impact at low temperatures. Meanwhile, most of the other eigenvalues form nearly symmetric pairs. We note that it still has a nearly degenerate ground state in the weak field regime (see Appendix~\ref{appA} for more details). Therefore, this structure suggests that the energy spectrum is not highly sensitive to small variations in the magnetic field. The magnetic field term, \( h\sum_i \hat{\sigma}_i^x \), primarily shifts the energy levels, lifting degeneracies and modifying the spectrum. For instance, in Fig.~\ref{fig:Eb} (bottom panel), the higher degenerate levels that were observed at \( h = 0 \) are slightly split when the field is increased to \( h = 0.08 \) or $h=0.45$, indicating a moderate lifting of degeneracy due to the weak transverse field, and they are still clustered around the zero line.
However, due to the pairing symmetry, the energy gaps between levels remain largely unchanged with variations in $h$. Consequently, the high degree of spectral symmetry in the $C_4$ energy structure limits its sensitivity to $h$, leading to reduced precision in estimating $h$.

On the other hand, the eigenvalue spectrum of the pan graph and $K_4$, shown in Fig.~\ref{fig:Eb}(bottom panel), displays six lowest degenerate levels (ground state) at \( h = 0 \). However, when a small transverse field \( h = 0.45 \) is applied, all these six low-energy levels in these two graphs experience noticeable splitting due to degenerate perturbation effects. Unlike the \( C_4 \) configuration, these shifts are more substantial, indicating that the low-energy spectrum of the pan graph is highly sensitive to the magnetic field as compared to the $K_4$. Therefore, among the configurations considered, the pan graph exhibits the largest and sharpest spectrum changes with the magnetic field strength (see Fig.~\ref{fig:App2} for more details), indicating a strong sensitivity to $h$ in the low-energy sector.

Furthermore, the spectrum of the pan graph at \( h = 0.45 \) contains a single zero eigenvalue, indicating a unique high-energy state or a less degenerate excited subspace. The absence of pairing in the low-energy levels or ground state implies that the energy gaps are more susceptible to perturbations from the magnetic field. The increased sensitivity of the energy spectrum to $h$ arises because the energy levels are not constrained by pairing symmetries, allowing for more flexible shifts in response to changes in $h$ (see Fig.~\ref{fig:Eb}). As a result, the QFI is higher for the pan graph configuration. Here, absence of ground-state degeneracy refers to the spectrum at finite magnetic field  $h>0$; in the strict $h=0$ limit, the pan graph does exhibit degeneracy, as noted in Appendix~\ref{appA}. Similarly, the energy spectrum of $K_4$ also lacks ground-state degeneracy at finite field $h>0$, and hence it shows higher QFI (see QFI in Figs.~\ref{fig5}(a) and (b)). In summary, spectra that are more irregular and lack ground-state degeneracy exhibit higher sensitivity to magnetic-field perturbations, highlighting an important design principle for optimizing quantum-sensing performance.

\subsection{Antiferromagnetic thermometers}
\begin{figure}[t]
    \centering
         \subfloat[]{
        \includegraphics[scale=0.55]{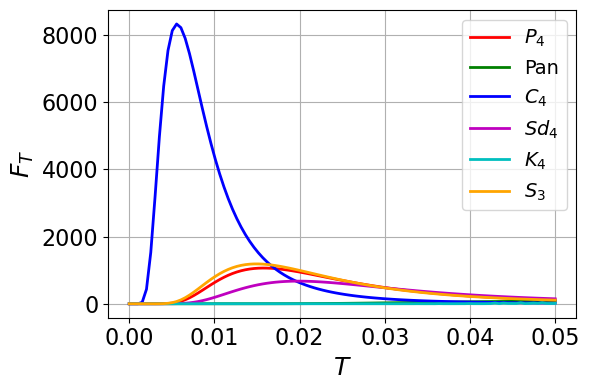}} \\
         \subfloat[]{
        \includegraphics[scale=0.55]{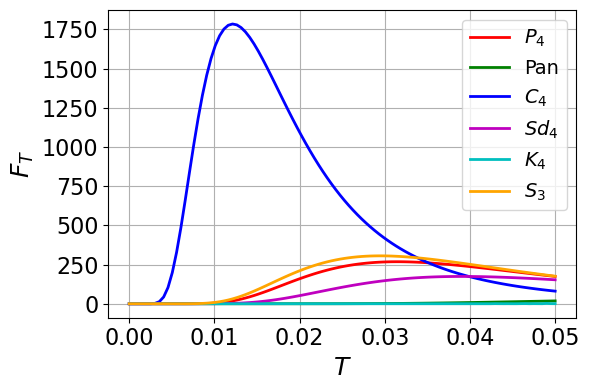}}     
        \caption{The QFI ($F_T$) for temperature estimation is plotted as a function of temperature \( T \) for two values of magnetic fields \( h = 0.45 \) (a) and \( h = 0.55 \) (b).  The different colors, such as red, green, blue, magenta, cyan, and orange, represent the configurations \( P_4 \), \( C_4 \), Pan, \( Sd_4 \), \( K_4 \), and \( S_3 \), respectively. All plots correspond to the antiferromagnetic coupling (\( J = -1 \)).}
    \label{fig6}
\end{figure}

\begin{figure}[t!]
    \centering
    \includegraphics[scale=0.68]{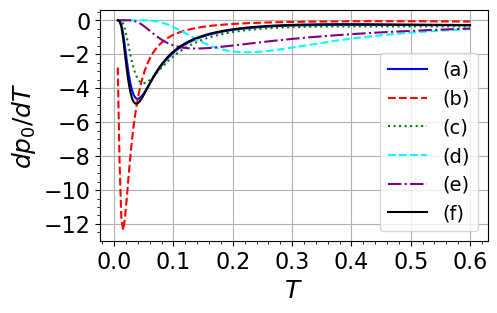}
    \caption{Rate of change of Boltzmann factor $dp_0/dT$ for the ground state as a function of temperature $T$. The different curves correspond to the configurations labeled (a), (b), (c), (d), and (e), illustrating their sensitivity to temperature changes. The rest of the parameters are set to $J=-1$ and $h=0.5$.}
    \label{fig:bolt}
\end{figure}

The QFI as a function of temperature T for various structures
of the quantum sensor is illustrated in Fig.~\ref{fig6} when antiferromagnetic coupling $J=-1$ is considered between the qubits. Our results reveal that the $C_4$ graph exhibits the highest
sensitivity for temperature estimation (blue curve in Fig.~\ref{fig6}(a)), while the precision achieved by other configurations is small. Furthermore, we observe that stronger magnetic fields have a detrimental effect on the precision of temperature estimation (see Fig.~\ref{fig6}(b) where $h=0.55$). It is interesting to note that the maximum interactions between the qubits are present in the
$K_4$ graph depicted in Fig.~\ref{fig2}(d). For this configuration, the temperature precision is the least compared to all other graphs, which shows that maximum connectivity does not help in temperature sensing for $J=-1$.
We examine the rate of the Boltzmann factor for each structure to determine why different configurations yield varying temperature precision. The Boltzmann factor governs the probability of a system being in a particular state at temperature $T$, such that $P(\sigma) \propto e^{-\beta \hat{H}(\sigma)}$, where $\hat{H}(\sigma)$ is the energy of a given spin configuration $\sigma$ and $\beta = 1/k_B T$. To analyze the system’s response to temperature changes, we compute the rate of change of the \textit{full Boltzmann distribution} with respect to temperature. For each eigenstate with energy $E_i$, the Boltzmann weight is given by
\begin{equation}
    p_j(T) = \frac{e^{-\beta E_j}}{\mathcal{Z}(T)},
\end{equation}
where $\beta = \frac{1}{k_B T}$ and $\mathcal{Z}(T) = \sum_j e^{-\beta E_j}$ is the partition function. 
In particular, we focus on the ground state population $p_0(T)$, where $E_0$ is the lowest energy eigenvalue. Temperature sensitivity may then be assessed by the derivative
\begin{equation}
\frac{d}{dT} p_0(T) = \frac{p_0(T)}{T^2} \left( E_0 - \bar{E} \right), \qquad \bar{E} = \sum_j p_j(T) E_j,
\end{equation}
which characterizes how the thermal population of the ground state changes with temperature.
This approach allows us to quantify the sensitivity of the ground state to temperature changes and compare the performance of different configurations in temperature sensing. In Fig.~\ref{fig:bolt}, we plot the rate of change of Boltzmann factor $dp_0(T)/dT$ as a function of bath temperature $T$ for different configurations. We can see that the configuration with the highest QFI (Fig.~\ref{fig2}(b)) corresponds to the one with the sharpest change in $dp_0(T)/d T$ values at low temperature, as can be seen by the red dashed curve in Fig.~\ref{fig:bolt}, indicating that it is the most sensitive to temperature changes and thus provides the highest precision in temperature estimation. However, when $J = -1$, the sharpest variation in the temperature derivative of the Boltzmann factor is observed for configuration $K_4$, in agreement with our earlier predictions shown in Fig.~\ref{fig4}. Moreover, the QFI maximum also lies in the same low-temperature regime. In contrast, the other configurations show relatively smooth and less pronounced variations, meaning their temperature sensitivity—and consequently their QFI—remains lower.

Degeneracy plays a crucial role in enhancing precision and expanding the measurable temperature
range in quantum thermometry~\cite{PhysRevLett.114.220405, campbell2018precision}. In our setup,
this enhancement arises from the presence of an \emph{exact} ground-state degeneracy in the
\(C_4\) configuration at finite magnetic field \(h>0\), as shown in Fig.~\ref{fig:Eb}(bottom panel).
The pairing of eigenvalues in the $C_4$ graph reflects a high degree of spectral symmetry, which includes ground-state degeneracy. Both the $C_4$ and $K_4$ graphs feature zero eigenvalues; however, these may not have a significant impact at low temperatures. By contrast, the pan and \(K_4\) graphs do not exhibit ground-state degeneracy at a finite field.
Their spectra are more irregular and lack the symmetry compared to \(C_4\).
Such spectra can enhance magnetic-field sensitivity, but do not provide the same temperature-sensing
advantages. These observations reinforce that the structure of the energy spectrum, particularly its graph and spectral symmetry and degeneracy, plays a critical role in determining the precision of temperature estimation.

From these observations, we conclude that compared to the other five structures/configurations investigated in this study, the $C_4$ structure demonstrates superior performance in terms of estimation precision of $T$, particularly in the low-temperature regime (when $J=-1$). While the other configurations exhibit varying degrees of sensitivity, none surpass the precision achieved by the $C_4$ structure. Note that qubit connectivity plays a crucial role in temperature sensing, where graph symmetry alone does not determine the optimal sensing configuration. While $K_4$ is highly symmetric, it is suboptimal for thermometry when $J=-1$. In contrast, $C_4$ has lower symmetry but performs optimally for temperature sensing. This highlights the importance of the interplay of the graph topology and spectral structure of the network Hamiltonian in optimizing quantum sensor performance for temperature estimation.

We can further note that the higher temperatures result in lower QFI values across all \( h \), highlighting the detrimental impact of thermal fluctuations on parameter estimation. These findings suggest that low-temperature regimes are optimal for magnetic field sensing with enhanced precision, particularly for weak fields. However, as \( T \) increases, thermal fluctuations dominate, leading to a loss of estimation precision of $h$.
Among all configurations considered, a pan graph and $K_4$ graph yield the highest QFI due to the absence of ground-state degeneracy, indicating their superior performance for more precise magnetic field estimation, especially in the low-temperature and weak-field regime.
\begin{figure}
    \centering
    \subfloat[]{
    \includegraphics[scale = 0.55]{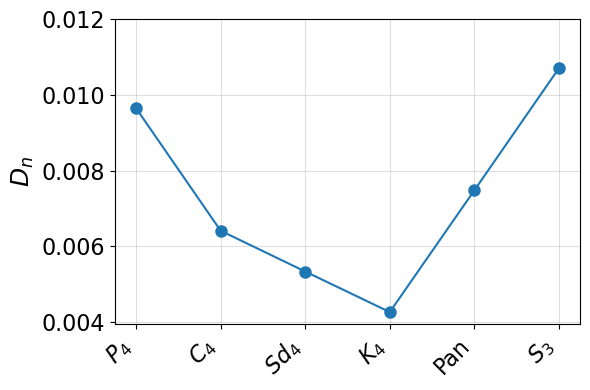}}\\
    \subfloat[]{
    \includegraphics[scale = 0.55]{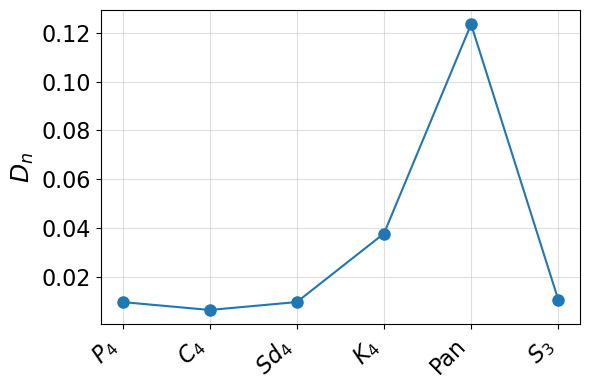}}
    \caption{Spectral deformation measure $D_n$(h) for all six connected graphs in the ferromagnetic $J=1$ (a) and antiferromagnetic $J=-1$ (b) coupling. The magnetic field is fixed at $h=0.08$. The index in the definition of $D_n$ is set at $n=1$ in both cases.}
    \label{fig:Dn}
\end{figure}
\section{Role of degeneracy and energy deformation}\label{deform}
In this section, we discuss how ground-state degeneracy and energy spectrum deformation can also serve as indicators of enhanced quantum sensitivity to magnetic fields. These quantifiers may provide valuable guidelines for identifying the graph structures that are most sensitive to the external magnetic field, especially in scenarios that involve many qubits, where direct calculation of the QFI becomes computationally expensive. Thus, analyzing degeneracy and energy deformation offers an efficient alternative to pinpoint configurations with optimal metrological performance without full QFI evaluation.
\subsection{Spectral sensitivity measure}
In quantum sensing applications, one needs to identify systems with enhanced sensitivity to small external perturbations. One approach is to study the response of the Hamiltonian spectrum under a weak perturbation parameter, such as the magnetic field $h$. For a system with 
Hamiltonian
\begin{equation}
    \hat{H}(h)=\hat{H}_0+h\hat{V},
\end{equation}
where $h$ is a small parameter and $\hat{V}$ is a perturbing observable, we may introduce a global spectral deformation measure  as
\begin{equation}
    D_n(h) = \sqrt{\sum_{i=0}^{n-1} \left[ E_i(h) - E_i(0) \right]^2 },
\end{equation}
where \(E_i(h)\) are the eigenvalues in the presence of a finite field \(h>0\), whereas \(E_i(0)\) correspond to the unperturbed spectrum at \(h=0\). The parameter \(n\) specifies how many of the lowest levels are included; for example, \(n=1\) means only the ground state, while \(n=2\) includes the ground and first excited state.  Thus, \(D_n(h)\) quantifies how strongly the lowest part of the spectrum deforms under the perturbation, and a larger \(D_n(h)\) typically correlates with a larger QFI for estimating \(h\). This measure is useful for comparing the sensitivity of different system configurations, such as spin networks defined on distinct graphs. 
 For a representative comparison, we plot the spectral deformation measure \(D_n\) for all six graphs in figure~\ref{fig:Dn} for the ground state ($n=1$). For ferromagnetic coupling \(J=1\), the star graph \(S_3\) exhibits the highest \(D_n\) value, followed by the linear chain \(P_4\). Interestingly, the QFI of the \(P_4\) graph is slightly larger than that of \(S_3\), indicating that \(D_n(h)\) does not fully capture the metrological performance in this regime.
 On the other hand, in the antiferromagnetic case \(J=-1\), the pan graph displays the highest \(D_n\) (see Fig.~\ref{fig:Dn}(b)), which corresponds to its highest QFI for the estimation of \(h\) (see Fig.~\ref{fig5}). This demonstrates that the spectral deformation measure \(D_n\) effectively captures the sensitivity of the graphs, particularly in the antiferromagnetic regime.

While $D_n(h)$ is not analytically equivalent to fidelity susceptibility, the Bures metric, or the QFI, we use it as an exploratory heuristic that provides qualitative insight into how strongly the low-energy spectrum responds to a small external field. Its appeal lies in its simplicity: $D_n(h)$ is easy to compute and often correlates well with metrological performance, especially in the antiferromagnetic regime. Accordingly, we treat $D_n(h)$ as a complementary spectral proxy rather than a predictive tool for optimizing magnetometric performance. In summary, while \(D_n(h)\) serves as a reliable predictor of metrological performance in the antiferromagnetic regime, its correlation with the QFI weakens in the ferromagnetic case.

\subsection{Degeneracy and sensitivity}

To understand the mechanism of degeneracy in sensitivity, let us perform a perturbative analysis of the QFI in the presence of ground-state degeneracy.
This analysis applies to the \emph{zero-temperature} limit ($T=0$), where the QFI is governed purely by ground-state structure; therefore, the degeneracy-enhanced sensitivity derived here does not contradict our finite-temperature results, where graphs with less spectral degeneracy exhibit a stronger response to the field.

We consider a $g$-fold degenerate ground state $\{|\psi_j^{(0)}\rangle\}_{j=1}^g$ of the unperturbed Hamiltonian $\hat{H}_0$, all sharing the eigenvalue $E_0$.
A small perturbation of the form $h\hat{V}$ couples these ground states to the excited states $|\psi_k^{(0)}\rangle$ of $\hat{H}_0$, potentially lifting the degeneracy. Let the system be initialized in a uniform superposition over the degenerate ground space
\begin{equation}
    |\psi^{(0)}\rangle = \frac{1}{\sqrt{g}} \sum_{j=1}^{g} |\psi^{(0)}_j\rangle.
\end{equation}
Applying time-independent non-degenerate perturbation theory, the first-order correction to each ground state due to \( V \) is
\begin{equation}
    |\delta\psi_j\rangle = h \sum_{k \notin \text{deg.}} \frac{\langle \psi^{(0)}_k | V | \psi^{(0)}_j \rangle}{E^{(0)}_k - E_0} |\psi^{(0)}_k\rangle.
\end{equation}
The perturbed ground state becomes
\begin{equation}
    |\psi_j(h)\rangle = |\psi^{(0)}_j\rangle + |\delta\psi_j\rangle + \mathcal{O}(h^2).
\end{equation}
The QFI for estimating \( h \) in the case of a pure state \( |\psi(h)\rangle = \frac{1}{\sqrt{g}} \sum_{j=1}^{g} |\psi_j(h)\rangle \) is given by
\begin{equation}
    F_Q(h) = 4 \left( \langle \partial_h\psi(h) | \partial_h\psi(h) \rangle - |\langle \psi(h) |\partial_h\psi(h) \rangle|^2 \right),
\end{equation}
where $\partial_h$ denotes the partial derivative with respect to \( h \).
To leading order in \( h \), we can obtain
\begin{equation}
    F_Q(h) = 4h^2 \left[ \frac{1}{g} \sum_{j=1}^{g} \sum_{k \notin \text{deg.}} \left| \frac{\langle \psi^{(0)}_k | V | \psi^{(0)}_j \rangle}{E^{(0)}_k - E_0} \right|^2 \right] + \mathcal{O}(h^3).
\end{equation}
This shows explicitly that the presence of ground-state degeneracy \( g \) can enhance the QFI in the small \( h \) regime. Similarly, for a maximally mixed initial state within the degenerate ground state subspace, the QFI is approximately given by
\begin{equation}
    F_Q(h) \approx \frac{4h^2}{g} \sum_{i < j} \frac{|\langle \psi_i^{(0)} | \hat{V} | \psi_j^{(0)} \rangle|^2}{(E_i - E_j)^2},
\end{equation}
where \( g \) is the ground-state degeneracy and \( \hat{V} = \sum_i \hat{\sigma}_i^x \) is the perturbation operator. 
Although the prefactor \( 1/g \) appears due to the mixed-state assumption, the number of contributing off-diagonal terms grows as \( g(g - 1)/2 \). 
This can result in an overall enhancement of the QFI with increasing degeneracy, provided the energy differences \( E_i - E_j \) are not too large. 
\section{Conclusion}\label{conc}
In this work, we have studied the performance of a four-qubit quantum sensor governed by the transverse Ising model, considering all connected graph configurations for applications in magnetometry and thermometry. Our analysis includes both ferromagnetic and antiferromagnetic couplings, assessing their impact on the sensor’s behavior across different geometries. Using the QFI to quantify precision limits, we have analyzed in detail how the qubit arrangement and the spectral structure of the Hamiltonian determine the sensor’s metrological capabilities.

In the ferromagnetic regime and weak magnetic field, graphs with minimal connectivity, like the 
linear-chain $P_4$, provide optimal performance due to degenerate ground states and efficient spin flips. The star graph ($S_3$) exhibits similar behavior. Conversely, highly connected graphs, such 
as the complete graph $K_4$, require stronger fields to show significant sensitivity. For the estimation of temperature, more connected graphs like $K_4$ and $Sd_4$ perform better at weaker fields due to their stiffer structure and larger energy gaps. This indicates that low-energy-excitation-favoring structures perform well at low temperatures, while stiffer graphs maintain sensitivity in strong fields.

In the antiferromagnetic case, we have identified the triangular configuration with a pendant 
vortex (termed pan graph) as the most effective for magnetic field estimation, while the 
$C_4$ configuration is better for temperature sensing. We have discussed the differences in magnetic sensing precision across various configurations by analyzing the spectra of their eigenvalues. 
The $C_4$ graph configuration, which exhibits lower sensitivity to magnetic fields, has four exactly zero eigenvalues, with two additional eigenvalues lying very close to zero. This leads to a high degree of degeneracy, including nearly degenerate ground states. Such a spectrum limits the system’s responsiveness to magnetic field perturbations due to degenerate values, thereby lowering the QFI for magnetic field estimation. In contrast, the pan graph configuration, identified as optimal for magnetometry, 
exhibits a non-degenerate spectrum with only one zero eigenvalue and no pairing symmetries. 
The absence of degenerate ground states and low symmetry in the spectrum or pairing 
allows the energy levels to shift more distinctly under external field perturbations, 
enhancing the system’s sensitivity to the external field $h$. These energy spectrum features of the pan and $K_4$ graph configurations make them well-suited for the enhanced precision of magnetic field sensing, while graphs with degenerate energy levels yield the highest QFI for temperature estimation.

Furthermore, the configuration with the highest QFI for temperature estimation exhibits the sharpest variation in the rate of change of the Boltzmann factor as a function of temperature, which further confirms its superior sensitivity to temperature changes. The rapid change in occupation probabilities of energy states in this configuration enhances its response to temperature variations, making it optimal for quantum thermometry. In contrast, other configurations display more gradual variations, leading to lower temperature sensitivity and, hence, a lower QFI. The enhanced temperature sensitivity of the 
$C_4$ graph configuration is due to the degenerate ground states. Our results indicate that low temperatures enhance magnetic field sensing, whereas weak magnetic fields improve temperature 
estimation. Conversely, strong magnetic fields degrade QFI as a function of temperature.

Our analysis shows that the spectral sensitivity measure $D_n(h)$ provides 
useful information about how strongly a system responds to small perturbations, 
and it may serve as a practical tool for identifying graph configurations with 
enhanced magnetic-field sensitivity, especially when exploring larger networks 
where full QFI calculations become computationally demanding. This complementary understanding can be useful for designing graph-based quantum sensors with optimal precision. Our results provide insights into the role of graph connectivity in quantum sensing and highlight how configurational optimization can enhance measurement precision in a multi-qubit sensor. A systematic approach to generalizing our case study to a larger number of qubits or quantum sensor networks may be possible using graph theory. The symmetry of graphs can be analyzed in terms of their automorphism group size, which quantifies the number of ways to relabel the vertices while preserving the adjacency matrix. However, graph symmetry alone does not determine the optimal configuration for sensing. For example, the complete graph $K_4$ is not optimal for thermometry despite its high symmetry, as indicated by its large automorphism group of size $24$. This suggests that higher symmetry does not necessarily imply better sensitivity for parameter estimation tasks like temperature sensing.

\section*{Acknowledgement}
This work is supported by the Scientific and Technological Research Council 
(TÜBİTAK) of T\"urkiye under Project Grant No. 123F150. {MGAP acknowledges partial support from MUR - NextGenerationEU through Projects G53D23001110006-RISQUE, G53D23006270001-QWEST, and J13C22000680006-QMORE.}

\appendix
\begin{widetext} 
\section{Effect of external magnetic field on energy spectrum}\label{appA}
In this Appendix, we examine the eigenvalue spectra for six distinct configurations of the quantum sensor as a function of magnetic field $h$, for an antiferromagnetic coupling ($J=-1$). In Fig.~\ref{fig:Eb} (bottom panel for $J=-1$), we first investigate the case when no external magnetic field is applied ($h=0$). It is immediately apparent that, except for the \(K_4\) and pan graphs, all other configurations exhibit $2$-fold degenerate ground states. Similarly, the square configuration shows the highest degeneracy in the first excited state. When an external magnetic field is applied, it acts as a perturbation that does not completely lift the degeneracy of $C_4$ structure, only slightly shifting the energy levels. However, for a strong magnetic field of $h=0.45$, the ground-state degeneracy is lifted, and also the low-energy states do not remain degenerate, but this does not benefit the estimation of $h$. The \(C_4\) configuration, in particular, tries to maintain symmetry in its energy spectrum, resulting in minimal response to magnetic perturbations, making it a poor candidate for magnetic field sensing. However, it proves to be an excellent probe for temperature sensing, where degeneracy typically enhances measurement precision.
\begin{figure*}[t!]
    \centering
    \begin{tabular}{ccc}
        \includegraphics[scale=0.55]{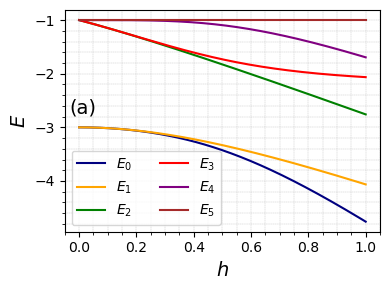} &
        \includegraphics[scale=0.55]{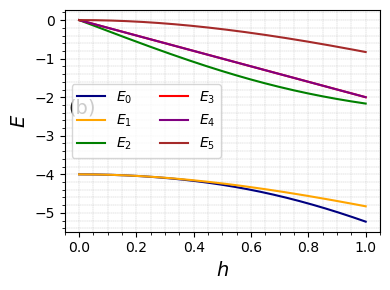} &
        \includegraphics[scale=0.55]{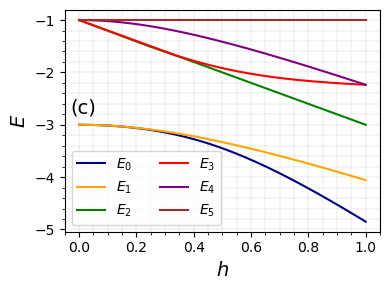} \\
        \includegraphics[scale=0.55]{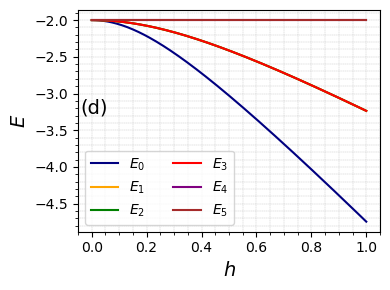} &
        \includegraphics[scale=0.55]{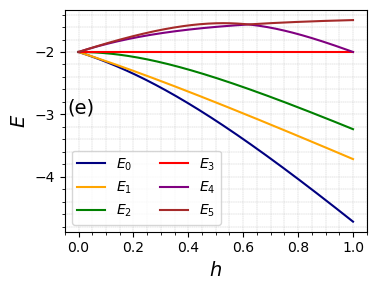} &
        \includegraphics[scale=0.55]{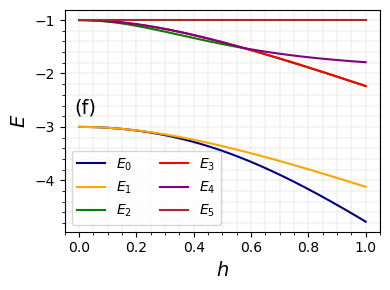}
    \end{tabular}
    \caption{The first six eigenvalues plotted as a function of external magnetic field $h$ for all six different graphs ($N=4$). The coupling strength is fixed at $J=1$. The panel labels (a) to (f) in these plots correspond to six different configurations of the $4$-qubit quantum sensor as shown in Fig.~\ref{fig2}.}
    \label{fig:App2}
\end{figure*}
On the other hand, the \(K_4\) graph and pan graph both exhibit six degenerate ground states when \(h = 0\) (see Fig.~\ref{fig:Eb} of the main text). Upon applying an external magnetic field, it acts as a perturbation and lifts the degeneracy. At \(h = 0.45\), all six energy levels are shifted due to degenerate perturbations. Specifically, for the \(K_4\) graph, the six degenerate states shift slightly under \(h = 0.45\), while for the pan graph, all six low-energy levels remain relevant due to greater shift in energy levels, and it shows the most significant change among the configurations under the same external field.

Fig.~\ref{fig:App2} shows the behavior of the first six eigenvalues as a function of the external magnetic field \(h\) for $J=-1$. From these plots, it is evident that all graphs, except for the \(K_4\) and pan graphs, exhibit nearly degenerate ground states in the weak field regime. This degeneracy is progressively lifted as the magnetic field strength increases, which explains the decrease in temperature precision in the strong magnetic field regime. In contrast, both the \(K_4\) and pan graphs show a more pronounced shift in the energy levels with no signs of degeneracy. As a result, these two graphs are most responsive to the external magnetic field, making them optimal configurations for magnetic field sensing.

\section{Effect of temperature on the estimation of the magnetic field}\label{effectT}

In this section, we present our results on the effect of temperature on the estimation of the parameter \( h \) in Fig.~\ref{fig:effectT}. We analyze the QFI for different temperature values for ferromagnetic ( $J = 1$) and antiferromagnetic ( $J = -1$ ) sensors of the magnetic field. Our findings show that increasing temperature drastically reduces the QFI, indicating a significant decline in estimation precision at higher temperatures for both the ferromagnetic ($J=1$) and antiferromagnetic ($J=-1$) interaction regimes. Notably, ferromagnetic sensors still maintain better sensitivity compared to antiferromagnetic sensors at small and intermediate temperatures. In particular, for $J = 1$, the QFI for the graphs $P_4$ and $S_3$ becomes nearly identical at higher temperatures, such as $T = 1.0$ and $T = 2.0$, indeed with the reduced precision. Similarly, for $J = -1$, increasing $T$ causes the QFI for the $P_4$ and $K_4$ graphs to converge. At sufficiently higher temperatures, the QFI for all the graphs converges. These observations indicate that higher temperatures degrade the metrological performance of all graph topologies, leading to similar precision levels across different graphs.

\begin{figure}[t!]
    \centering
     \begin{tabular}{ccc}
    \includegraphics[width=0.33\linewidth]{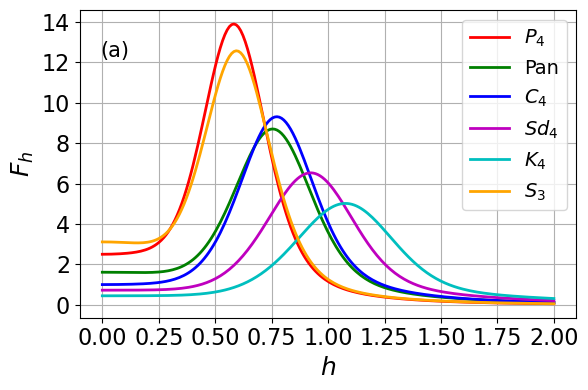}&
    \includegraphics[width=0.33\linewidth]{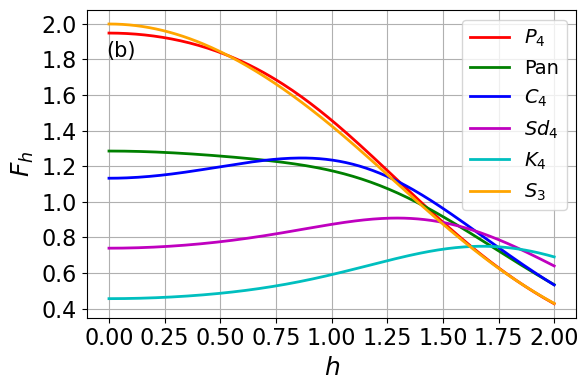}&
    \includegraphics[width=0.33\linewidth]{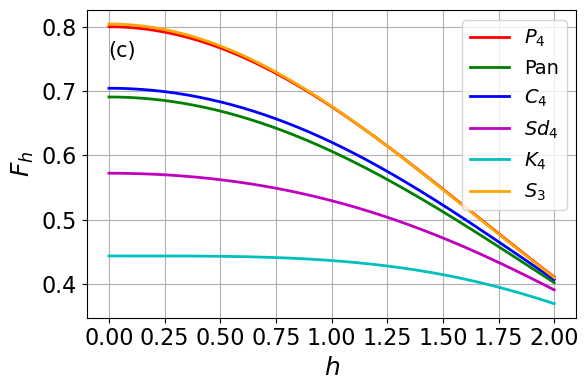}\\
     \includegraphics[width=0.33\linewidth]{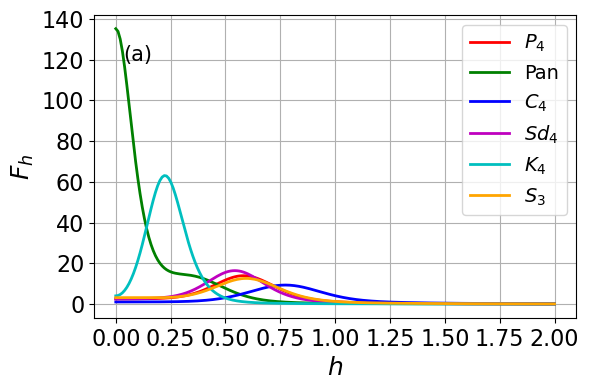}&
    \includegraphics[width=0.33\linewidth]{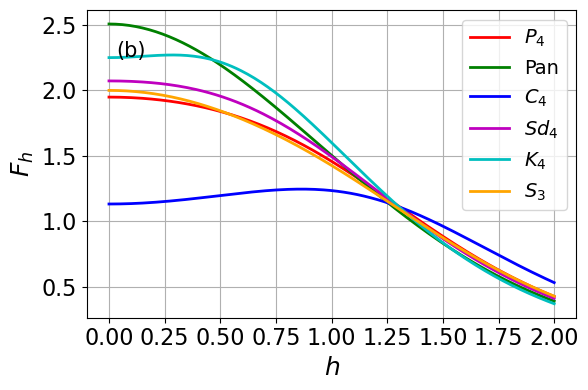}&
    \includegraphics[width=0.33\linewidth]{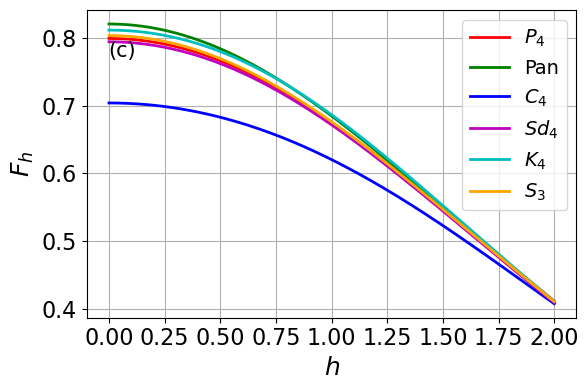}
        \end{tabular}
    \caption{\textbf{Top panel:} QFI $F_h$ as a function of magnetic field $h$ for different values of temperature $T$ when $J=1$. \textbf{Bottom panel:} QFI $F_h$ as a function of magnetic field $h$ for different values of temperature $T$ when $J=-1$. Here, the legends in each plot correspond to (a) $T=0.1$, (b) $T=1.0$, and (c) $T=2.0$, respectively.}
    \label{fig:effectT}
\end{figure}
\section{Effect of magnetic field on the estimation of the temperature}\label{effecth}
\begin{figure}[t!]
    \centering
     \begin{tabular}{ccc}
    \includegraphics[width=0.32\linewidth]{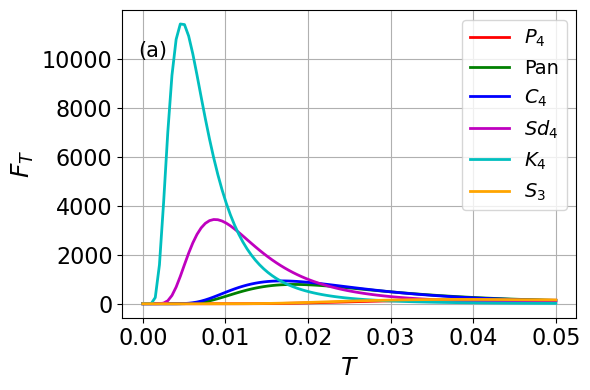}&
    \includegraphics[width=0.32\linewidth]{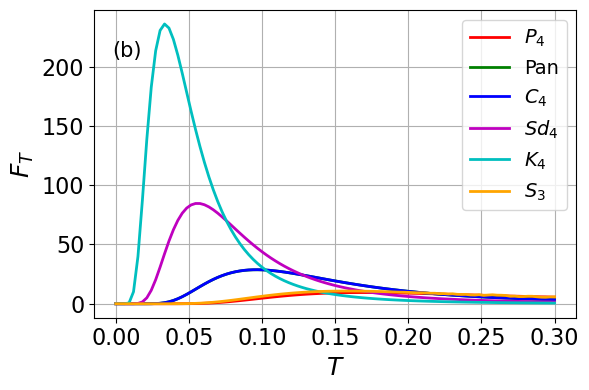}&
    \includegraphics[width=0.32\linewidth]{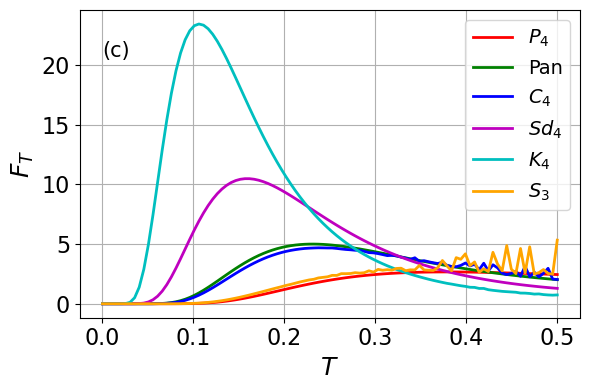}\\
     \includegraphics[width=0.32\linewidth]{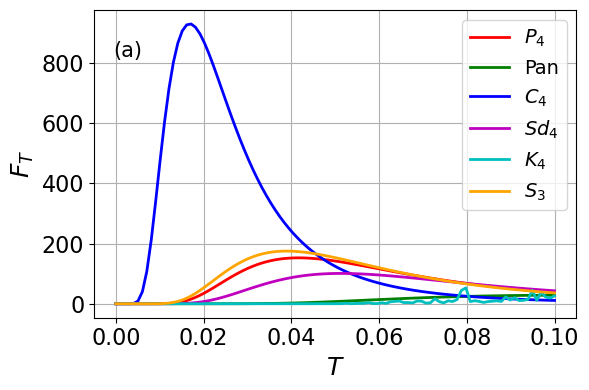}&
    \includegraphics[width=0.32\linewidth]{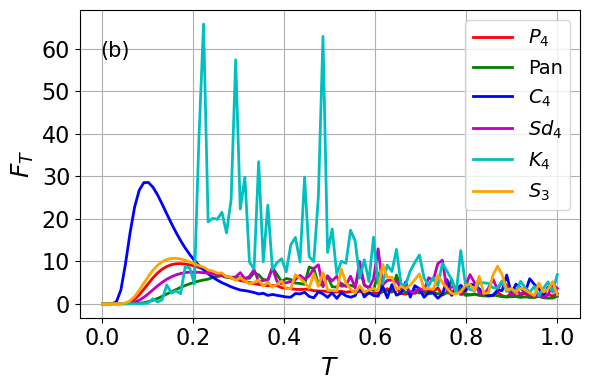}&
    \includegraphics[width=0.32\linewidth]{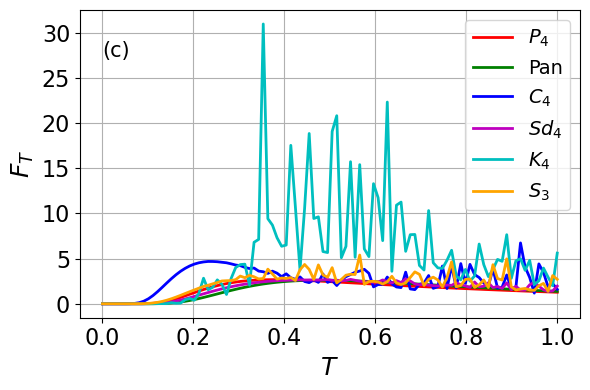}
        \end{tabular}
    \caption{\textbf{Top panel:} QFI $F_T$ as a function of temperature $T$ for different values of magnetic field $h$ when $J=1$. \textbf{Bottom panel:} QFI $F_T$ as a function of temperature $T$ for different values of magnetic field $h$ when $J=-1$. Here, the legends in each plot correspond to (a) $h=0.6$, (b) $h=1.0$, and (c) $h=1.4$, respectively.}
    \label{fig:effecth}
\end{figure}
Figure~\ref{fig:effecth} shows the QFI $F_T$ for temperature estimation across different magnetic fields $h$ for both ferromagnetic ($J = 1$, top panel) and antiferromagnetic ($J = -1$, bottom panel) couplings. For $J = 1$, at small $h$, the QFI for the complete graph $K_4$ exhibits a single dominant peak at low $T$, indicating an optimal range for high-precision thermometry. Increasing $h$ shifts this peak toward higher $T$ and reduces its height, reflecting the suppression of thermal sensitivity due to stronger field values. For $J = -1$, the cycle graph $C_4$ shows maximum QFI when the magnetic field is weak ($h=0.6$). However, a stronger magnetic field strength ($h=1.0$ and $h=1.4$) lowers the peak height of $C_4$ but also enhances oscillations in $K_4$ as $h$ increases. In the antiferromagnetic regime, frustration combined with the transverse field amplifies field-induced level mixing, leading to oscillatory QFI behavior in $K_4$. In addition, the peak QFI values for the $K_4$ graph exceed those observed for the $C_4$ graph. Overall, these results indicate that strong magnetic fields deteriorate the metrological performance of all graph topologies for temperature estimation.

\end{widetext}

\newpage
\bibliography{grp.bib}

@article{PhysRevA.90.022111,
  title = {Quantum metrology in Lipkin-Meshkov-Glick critical systems},
  author = {Salvatori, Giulio and Mandarino, Antonio and Paris, Matteo G. A.},
  journal = {Phys. Rev. A},
  volume = {90},
  issue = {2},
  pages = {022111},
  numpages = {11},
  year = {2014},
  month = {Aug},
  publisher = {American Physical Society},
  doi = {10.1103/PhysRevA.90.022111},
  url = {https://link.aps.org/doi/10.1103/PhysRevA.90.022111}
}

@article{PhysRevApplied.17.034073,
  title = {Non-Markovian Quantum Thermometry},
  author = {Zhang, Ning and Chen, Chong and Bai, Si-Yuan and Wu, Wei and An, Jun-Hong},
  journal = {Phys. Rev. Appl.},
  volume = {17},
  issue = {3},
  pages = {034073},
  numpages = {10},
  year = {2022},
  month = {Mar},
  publisher = {American Physical Society},
  doi = {10.1103/PhysRevApplied.17.034073},
  url = {https://link.aps.org/doi/10.1103/PhysRevApplied.17.034073}
}

@article{sidhu,
    author = {Sidhu, Jasminder S. and Kok, Pieter},
    title = {Geometric perspective on quantum parameter estimation},
    journal = {AVS Quantum Sci.},
    volume = {2},
    number = {1},
    pages = {014701},
    year = {2020},
    month = {02},
    issn = {2639-0213},
    doi = {10.1116/1.5119961},
    url = {https://doi.org/10.1116/1.5119961}
}

@article{PhysRevA.78.042105,
  title = {Quantum criticality as a resource for quantum estimation},
  author = {Zanardi, Paolo and Paris, Matteo G. A. and Campos Venuti, Lorenzo},
  journal = {Phys. Rev. A},
  volume = {78},
  issue = {4},
  pages = {042105},
  numpages = {7},
  year = {2008},
  month = {Oct},
  publisher = {American Physical Society},
  doi = {10.1103/PhysRevA.78.042105},
  url = {https://link.aps.org/doi/10.1103/PhysRevA.78.042105}
}

@article{PhysRevLett.120.080501,
  title = {Multiparameter Estimation in Networked Quantum Sensors},
  author = {Proctor, Timothy J. and Knott, Paul A. and Dunningham, Jacob A.},
  journal = {Phys. Rev. Lett.},
  volume = {120},
  issue = {8},
  pages = {080501},
  numpages = {6},
  year = {2018},
  month = {Feb},
  publisher = {American Physical Society},
  doi = {10.1103/PhysRevLett.120.080501},
  url = {https://link.aps.org/doi/10.1103/PhysRevLett.120.080501}
}

@article{PhysRevResearch.3.033011,
  title = {Protocols for estimating multiple functions with quantum sensor networks: Geometry and performance},
  author = {Bringewatt, Jacob and Boettcher, Igor and Niroula, Pradeep and Bienias, Przemyslaw and Gorshkov, Alexey V.},
  journal = {Phys. Rev. Res.},
  volume = {3},
  issue = {3},
  pages = {033011},
  numpages = {15},
  year = {2021},
  month = {Jul},
  publisher = {American Physical Society},
  doi = {10.1103/PhysRevResearch.3.033011},
  url = {https://link.aps.org/doi/10.1103/PhysRevResearch.3.033011}
}

@article{PhysRevA.111.052216,
  title = {Topological quantum thermometry},
  author = {Srivastava, Anubhav Kumar and Bhattacharya, Utso and Lewenstein, Maciej and P\l{}odzie\ifmmode \acute{n}\else \'{n}\fi{}, Marcin},
  journal = {Phys. Rev. A},
  volume = {111},
  issue = {5},
  pages = {052216},
  numpages = {9},
  year = {2025},
  month = {May},
  publisher = {American Physical Society},
  doi = {10.1103/PhysRevA.111.052216},
  url = {https://link.aps.org/doi/10.1103/PhysRevA.111.052216}
}

@article{PhysRevA.109.032608,
  title = {Quantum magnetometry using discrete-time quantum walk},
  author = {Shukla, Kunal and Chandrashekar, C. M.},
  journal = {Phys. Rev. A},
  volume = {109},
  issue = {3},
  pages = {032608},
  numpages = {11},
  year = {2024},
  month = {Mar},
  publisher = {American Physical Society},
  doi = {10.1103/PhysRevA.109.032608},
  url = {https://link.aps.org/doi/10.1103/PhysRevA.109.032608}
}

@article{Li18,
author = {Bei-Bei Li and Jan B\'{i}lek and Ulrich B. Hoff and Lars S. Madsen and Stefan Forstner and Varun Prakash and Clemens Sch\"{a}fermeier and Tobias Gehring and Warwick P. Bowen and Ulrik L. Andersen},
journal = {Optica},
number = {7},
pages = {850--856},
publisher = {Optica Publishing Group},
title = {Quantum enhanced optomechanical magnetometry},
volume = {5},
month = {Jul},
year = {2018},
url = {https://opg.optica.org/optica/abstract.cfm?URI=optica-5-7-850},
doi = {10.1364/OPTICA.5.000850}
}

@Article{Aslam2023,
author={Aslam, Nabeel
and Zhou, Hengyun
and Urbach, Elana K.
and Turner, Matthew J.
and Walsworth, Ronald L.
and Lukin, Mikhail D.
and Park, Hongkun},
title={Quantum sensors for biomedical applications},
journal={Nat Rev. Phys.},
year={2023},
month={Mar},
day={01},
volume={5},
number={3},
pages={157-169},
issn={2522-5820},
doi={10.1038/s42254-023-00558-3},
url={https://doi.org/10.1038/s42254-023-00558-3}
}

@article{Albarelli_2017,
doi = {10.1088/1367-2630/aa9840},
url = {https://dx.doi.org/10.1088/1367-2630/aa9840},
year = {2017},
month = {dec},
publisher = {IOP Publishing},
volume = {19},
number = {12},
pages = {123011},
author = {Albarelli, Francesco and Rossi, Matteo A C and Paris, Matteo G A and Genoni, Marco G},
title = {Ultimate limits for quantum magnetometry via time-continuous measurements},
journal = {New J. Phys.}
}

@article{rossi2015entangled,
  title={Entangled quantum probes for dynamical environmental noise},
  author={Rossi, Matteo AC and Paris, Matteo GA},
  journal={Physical Review A},
  volume={92},
  number={1},
  pages={010302},
  year={2015},
  publisher={APS}
}

@article{PhysRevLett.134.010801,
  title = {Fundamental Limits of Metrology at Thermal Equilibrium},
  author = {Abiuso, Paolo and Sekatski, Pavel and Calsamiglia, John and Perarnau-Llobet, Mart\'{\i}},
  journal = {Phys. Rev. Lett.},
  volume = {134},
  issue = {1},
  pages = {010801},
  numpages = {7},
  year = {2025},
  month = {Jan},
  publisher = {American Physical Society},
  doi = {10.1103/PhysRevLett.134.010801},
  url = {https://link.aps.org/doi/10.1103/PhysRevLett.134.010801}
}

@article{montenegro2024review,
title = {Review: Quantum metrology and sensing with many-body systems},
journal = {Phys. Rep.},
volume = {1134},
pages = {1-62},
year = {2025},
issn = {0370-1573},
doi = {https://doi.org/10.1016/j.physrep.2025.05.005},
url = {https://www.sciencedirect.com/science/article/pii/S0370157325001565},
author = {Victor Montenegro and Chiranjib Mukhopadhyay and Rozhin Yousefjani and Saubhik Sarkar and Utkarsh Mishra and Matteo G.A. Paris and Abolfazl Bayat}
}

@article{RevModPhys.89.035002,
  title = {Quantum sensing},
  author = {Degen, C. L. and Reinhard, F. and Cappellaro, P.},
  journal = {Rev. Mod. Phys.},
  volume = {89},
  issue = {3},
  pages = {035002},
  numpages = {39},
  year = {2017},
  month = {Jul},
  publisher = {American Physical Society},
  doi = {10.1103/RevModPhys.89.035002},
  url = {https://link.aps.org/doi/10.1103/RevModPhys.89.035002}
}

@article{TAYLOR20161,
title = {Quantum metrology and its application in biology},
journal = {Phys. Rep.},
volume = {615},
pages = {1-59},
year = {2016},
note = {Quantum metrology and its application in biology},
issn = {0370-1573},
doi = {https://doi.org/10.1016/j.physrep.2015.12.002},
url = {https://www.sciencedirect.com/science/article/pii/S0370157315005001},
author = {Michael A. Taylor and Warwick P. Bowen}
}

@article{Wang_2018,
doi = {10.1088/1367-2630/aab03a},
url = {https://dx.doi.org/10.1088/1367-2630/aab03a},
year = {2018},
month = {mar},
publisher = {IOP Publishing},
volume = {20},
number = {3},
pages = {033034},
author = {Wang, Zhihai and Wu, Wei and Cui, Guodong and Wang, Jin},
title = {Coherence enhanced quantum metrology in a nonequilibrium optical molecule},
journal = {New J. Phys.}
}

@article{PhysRevA.94.012101,
  title = {Usefulness of entanglement-assisted quantum metrology},
  author = {Huang, Zixin and Macchiavello, Chiara and Maccone, Lorenzo},
  journal = {Phys. Rev. A},
  volume = {94},
  issue = {1},
  pages = {012101},
  numpages = {5},
  year = {2016},
  month = {Jul},
  publisher = {American Physical Society},
  doi = {10.1103/PhysRevA.94.012101},
  url = {https://link.aps.org/doi/10.1103/PhysRevA.94.012101}
}

@Article{DeMille2024,
author={DeMille, David
and Hutzler, Nicholas R.
and Rey, Ana Maria
and Zelevinsky, Tanya},
title={Quantum sensing and metrology for fundamental physics with molecules},
journal={Nat. Phys.},
year={2024},
month={May},
day={01},
volume={20},
number={5},
pages={741-749},
issn={1745-2481},
doi={10.1038/s41567-024-02499-9},
url={https://doi.org/10.1038/s41567-024-02499-9}
}

@article{Dedyulin_2022,
doi = {10.1088/1361-6501/ac75b1},
url = {https://dx.doi.org/10.1088/1361-6501/ac75b1},
year = {2022},
month = {jun},
publisher = {IOP Publishing},
volume = {33},
number = {9},
pages = {092001},
author = {Dedyulin, S and Ahmed, Z and Machin, G},
title = {Emerging technologies in the field of thermometry},
journal = {Meas, Sci. Technol.}
}

@article{PhysRevApplied.13.054057,
  title = {Practical Applications of Quantum Sensing: A Simple Method to Enhance the Sensitivity of Nitrogen-Vacancy-Based Temperature Sensors},
  author = {Moreva, E. and Bernardi, E. and Traina, P. and Sosso, A. and Tchernij, S. Ditalia and Forneris, J. and Picollo, F. and Brida, G. and Pastuovi\ifmmode \acute{c}\else \'{c}\fi{}, \ifmmode \check{Z}\else \v{Z}\fi{}. and Degiovanni, I. P. and Olivero, P. and Genovese, M.},
  journal = {Phys. Rev. Appl.},
  volume = {13},
  issue = {5},
  pages = {054057},
  numpages = {9},
  year = {2020},
  month = {May},
  publisher = {American Physical Society},
  doi = {10.1103/PhysRevApplied.13.054057},
  url = {https://link.aps.org/doi/10.1103/PhysRevApplied.13.054057}
}

@article{PhysRevResearch.2.033389,
  title = {Quantum sensing of open systems: Estimation of damping constants and temperature},
  author = {Wang, J. and Davidovich, L. and Agarwal, G. S.},
  journal = {Phys. Rev. Res.},
  volume = {2},
  issue = {3},
  pages = {033389},
  numpages = {8},
  year = {2020},
  month = {Sep},
  publisher = {American Physical Society},
  doi = {10.1103/PhysRevResearch.2.033389},
  url = {https://link.aps.org/doi/10.1103/PhysRevResearch.2.033389}
}

@article{PhysRevA.108.032220,
  title = {Multispin probes for thermometry in the strong-coupling regime},
  author = {Brenes, Marlon and Segal, Dvira},
  journal = {Phys. Rev. A},
  volume = {108},
  issue = {3},
  pages = {032220},
  numpages = {11},
  year = {2023},
  month = {Sep},
  publisher = {American Physical Society},
  doi = {10.1103/PhysRevA.108.032220},
  url = {https://link.aps.org/doi/10.1103/PhysRevA.108.032220}
}

@Article{Gottscholl2021,
author={Gottscholl, Andreas
and Diez, Matthias
and Soltamov, Victor
and Kasper, Christian
and Krau{\ss}e, Dominik
and Sperlich, Andreas
and Kianinia, Mehran
and Bradac, Carlo
and Aharonovich, Igor
and Dyakonov, Vladimir},
title={Spin defects in hBN as promising temperature, pressure and magnetic field quantum sensors},
journal={Nat. Commun.},
year={2021},
month={Jul},
day={22},
volume={12},
number={1},
pages={4480},
issn={2041-1723},
doi={10.1038/s41467-021-24725-1},
url={https://doi.org/10.1038/s41467-021-24725-1}
}

@article{KLIMOV2018308,
title = {Towards replacing resistance thermometry with photonic thermometry},
journal = {Sensors Actuat. A: Phys.},
volume = {269},
pages = {308-312},
year = {2018},
issn = {0924-4247},
doi = {https://doi.org/10.1016/j.sna.2017.11.055},
url = {https://www.sciencedirect.com/science/article/pii/S0924424717318940},
author = {Nikolai Klimov and Thomas Purdy and Zeeshan Ahmed}
}

@article{Li,
    author = {Li, Bei-Bei and Wang, Qing-Yan and Xiao, Yun-Feng and Jiang, Xue-Feng and Li, Yan and Xiao, Lixin and Gong, Qihuang},
    title = {On chip, high-sensitivity thermal sensor based on high-Q polydimethylsiloxane-coated microresonator},
    journal = {Appl. Phys. Lett.},
    volume = {96},
    number = {25},
    pages = {251109},
    year = {2010},
    month = {06},
    issn = {0003-6951},
    doi = {10.1063/1.3457444},
    url = {https://doi.org/10.1063/1.3457444}
}

@article{dong,
    author = {Dong, C.-H. and He, L. and Xiao, Y.-F. and Gaddam, V. R. and Ozdemir, S. K. and Han, Z.-F. and Guo, G.-C. and Yang, L.},
    title = {Fabrication of high-Q polydimethylsiloxane optical microspheres for thermal sensing},
    journal = {Appl. Phys. Lett.},
    volume = {94},
    number = {23},
    pages = {231119},
    year = {2009},
    month = {06},
    issn = {0003-6951},
    doi = {10.1063/1.3152791},
    url = {https://doi.org/10.1063/1.3152791}
}

@article{Purdy,
author = {T. P. Purdy  and K. E. Grutter  and K. Srinivasan  and J. M. Taylor },
title = {Quantum correlations from a room-temperature optomechanical cavity},
journal = {Science},
volume = {356},
number = {6344},
pages = {1265-1268},
year = {2017},
doi = {10.1126/science.aag1407},
URL = {https://www.science.org/doi/abs/10.1126/science.aag1407}
}

@article{Ishizaki_2012,
  author = {Ishizaki, Akihito and Fleming, Graham R.},
  title = {Quantum Coherence in Photosynthetic Light Harvesting}, 
  journal = {Ann. Rev. Condens. Matter Phys.},
  year = {2012},
  volume = {3},
  pages = {333--361},
  doi = {10.1146/annurev-conmatphys-020911-125126},
  url = {https://www.annualreviews.org/content/journals/10.1146/annurev-conmatphys-020911-125126},
  publisher = {Annual Reviews},
  issn = {1947-5462}
}

@article{PhysRevA.109.012424,
  title = {Geometrical optimization of spin clusters for the preservation of quantum coherence},
  author = {Gassab, Lea and Pusuluk, Onur and M\"ustecapl\ifmmode \imath \else \i \fi{}o\ifmmode \breve{g}\else \u{g}\fi{}lu, \"Ozg\"ur E.},
  journal = {Phys. Rev. A},
  volume = {109},
  issue = {1},
  pages = {012424},
  numpages = {8},
  year = {2024},
  month = {Jan},
  publisher = {American Physical Society},
  doi = {10.1103/PhysRevA.109.012424},
  url = {https://link.aps.org/doi/10.1103/PhysRevA.109.012424}
}

@article{Upadhyay_2024,
doi = {10.1088/1742-5468/ad8f2c},
url = {https://dx.doi.org/10.1088/1742-5468/ad8f2c},
year = {2024},
month = {nov},
publisher = {IOP Publishing},
volume = {2024},
number = {11},
pages = {113104},
author = {Upadhyay, Vipul and Marathe, Rahul},
title = {Current circulation near additional energy degeneracy points in quadratic Fermionic networks},
journal = {J. Stat. Mech.},
}

@article{PhysRevE.104.014136,
  title = {Role of topology in determining the precision of a finite thermometer},
  author = {Candeloro, Alessandro and Razzoli, Luca and Bordone, Paolo and Paris, Matteo G. A.},
  journal = {Phys. Rev. E},
  volume = {104},
  issue = {1},
  pages = {014136},
  numpages = {14},
  year = {2021},
  month = {Jul},
  publisher = {American Physical Society},
  doi = {10.1103/PhysRevE.104.014136},
  url = {https://link.aps.org/doi/10.1103/PhysRevE.104.014136}
}

@article{PhysRevA.100.053825,
  title = {Quantum metrology enhanced by coherence-induced driving in a cavity-QED setup},
  author = {Cheng, Weijun and Hou, S. C. and Wang, Zhihai and Yi, X. X.},
  journal = {Phys. Rev. A},
  volume = {100},
  issue = {5},
  pages = {053825},
  numpages = {7},
  year = {2019},
  month = {Nov},
  publisher = {American Physical Society},
  doi = {10.1103/PhysRevA.100.053825},
  url = {https://link.aps.org/doi/10.1103/PhysRevA.100.053825}
}

@article{Chowdhury_2019,
doi = {10.1088/2058-9565/ab05f1},
url = {https://dx.doi.org/10.1088/2058-9565/ab05f1},
year = {2019},
month = {mar},
publisher = {IOP Publishing},
volume = {4},
number = {2},
pages = {024007},
author = {Chowdhury, A and Vezio, P and Bonaldi, M and Borrielli, A and Marino, F and Morana, B and Pandraud, G and Pontin, A and Prodi, G A and Sarro, P M and Serra, E and Marin, F},
title = {Calibrated quantum thermometry in cavity optomechanics},
journal = {Quantum Sci. Technol.}
}

@article{PhysRevLett.125.120603,
  title = {Detecting Acoustic Blackbody Radiation with an Optomechanical Antenna},
  author = {Singh, Robinjeet and Purdy, Thomas P.},
  journal = {Phys. Rev. Lett.},
  volume = {125},
  issue = {12},
  pages = {120603},
  numpages = {6},
  year = {2020},
  month = {Sep},
  publisher = {American Physical Society},
  doi = {10.1103/PhysRevLett.125.120603},
  url = {https://link.aps.org/doi/10.1103/PhysRevLett.125.120603}
}

@article{Aybar2022criticalquantum,
  doi = {10.22331/q-2022-09-19-808},
  url = {https://doi.org/10.22331/q-2022-09-19-808},
  title = {Critical quantum thermometry and its feasibility in spin systems},
  author = {Aybar, Enes and Niezgoda, Artur and Mirkhalaf, Safoura S. and Mitchell, Morgan W. and Benedicto Orenes, Daniel and Witkowska, Emilia},
  journal = {{Quantum}},
  issn = {2521-327X},
  publisher = {{Verein zur F{\"{o}}rderung des Open Access Publizierens in den Quantenwissenschaften}},
  volume = {6},
  pages = {808},
  month = sep,
  year = {2022}
}

@article{Galinskiy,
author = {I. Galinskiy and Y. Tsaturyan and M. Parniak and E. S. Polzik},
journal = {Optica},
number = {6},
pages = {718--725},
publisher = {Optica Publishing Group},
title = {Phonon counting thermometry of an ultracoherent membrane resonator near its motional ground state},
volume = {7},
month = {Jun},
year = {2020},
url = {https://opg.optica.org/optica/abstract.cfm?URI=optica-7-6-718},
doi = {10.1364/OPTICA.390939}
}

@article{PhysRevResearch.6.043094,
  title = {Criticality-enhanced precision in phase thermometry},
  author = {Yu, Mei and Nguyen, H. Chau and Nimmrichter, Stefan},
  journal = {Phys. Rev. Res.},
  volume = {6},
  issue = {4},
  pages = {043094},
  numpages = {13},
  year = {2024},
  month = {Nov},
  publisher = {American Physical Society},
  doi = {10.1103/PhysRevResearch.6.043094},
  url = {https://link.aps.org/doi/10.1103/PhysRevResearch.6.043094}
}

@article{PhysRevLett.112.110403,
  title = {Thermometry of Cold Atoms in Optical Lattices via Artificial Gauge Fields},
  author = {Roscilde, Tommaso},
  journal = {Phys. Rev. Lett.},
  volume = {112},
  issue = {11},
  pages = {110403},
  numpages = {5},
  year = {2014},
  month = {Mar},
  publisher = {American Physical Society},
  doi = {10.1103/PhysRevLett.112.110403},
  url = {https://link.aps.org/doi/10.1103/PhysRevLett.112.110403}
}

@article{Abiuso_2024,
doi = {10.1088/2058-9565/ad37d3},
url = {https://dx.doi.org/10.1088/2058-9565/ad37d3},
year = {2024},
month = {apr},
publisher = {IOP Publishing},
volume = {9},
number = {3},
pages = {035008},
author = {Abiuso, Paolo and Andrea Erdman, Paolo and Ronen, Michael and Noé, Frank and Haack, Géraldine and Perarnau-Llobet, Martí},
title = {Optimal thermometers with spin networks},
journal = {Quantum Sci. Technol.}
}

@article{PhysRevLett.107.083601,
  title = {Quantum Metrology with Entangled Coherent States},
  author = {Joo, Jaewoo and Munro, William J. and Spiller, Timothy P.},
  journal = {Phys. Rev. Lett.},
  volume = {107},
  issue = {8},
  pages = {083601},
  numpages = {4},
  year = {2011},
  month = {Aug},
  publisher = {American Physical Society},
  doi = {10.1103/PhysRevLett.107.083601},
  url = {https://link.aps.org/doi/10.1103/PhysRevLett.107.083601}
}

@article{PhysRevA.99.062330,
  title = {Lattice quantum magnetometry},
  author = {Razzoli, Luca and Ghirardi, Luca and Siloi, Ilaria and Bordone, Paolo and Paris, Matteo G. A.},
  journal = {Phys. Rev. A},
  volume = {99},
  issue = {6},
  pages = {062330},
  numpages = {9},
  year = {2019},
  month = {Jun},
  publisher = {American Physical Society},
  doi = {10.1103/PhysRevA.99.062330},
  url = {https://link.aps.org/doi/10.1103/PhysRevA.99.062330}
}

@article{PhysRevX.5.031010,
  title = {Improved Quantum Magnetometry beyond the Standard Quantum Limit},
  author = {Brask, J. B. and Chaves, R. and Ko\l{}ody\ifmmode \acute{n}\else \'{n}\fi{}ski, J.},
  journal = {Phys. Rev. X},
  volume = {5},
  issue = {3},
  pages = {031010},
  numpages = {13},
  year = {2015},
  month = {Jul},
  publisher = {American Physical Society},
  doi = {10.1103/PhysRevX.5.031010},
  url = {https://link.aps.org/doi/10.1103/PhysRevX.5.031010}
}

@article{PhysRevLett.127.113602,
  title = {In Situ Thermometry of Fermionic Cold-Atom Quantum Wires},
  author = {De Daniloff, Cl\'ement and Tharrault, Marin and Enesa, C\'edric and Salomon, Christophe and Chevy, Fr\'ed\'eric and Reimann, Thomas and Struck, Julian},
  journal = {Phys. Rev. Lett.},
  volume = {127},
  issue = {11},
  pages = {113602},
  numpages = {7},
  year = {2021},
  month = {Sep},
  publisher = {American Physical Society},
  doi = {10.1103/PhysRevLett.127.113602},
  url = {https://link.aps.org/doi/10.1103/PhysRevLett.127.113602}
}

@article{PhysRevResearch.4.043057,
  title = {QuanEstimation: An open-source toolkit for quantum parameter estimation},
  author = {Zhang, Mao and Yu, Huai-Ming and Yuan, Haidong and Wang, Xiaoguang and Demkowicz-Dobrza\ifmmode \acute{n}\else \'{n}\fi{}ski, Rafa\l{} and Liu, Jing},
  journal = {Phys. Rev. Res.},
  volume = {4},
  issue = {4},
  pages = {043057},
  numpages = {38},
  year = {2022},
  month = {Oct},
  publisher = {American Physical Society},
  doi = {10.1103/PhysRevResearch.4.043057},
  url = {https://link.aps.org/doi/10.1103/PhysRevResearch.4.043057}
}

@Article{Mok2021,
author={Mok, Wai-Keong
and Bharti, Kishor
and Kwek, Leong-Chuan
and Bayat, Abolfazl},
title={Optimal probes for global quantum thermometry},
journal={Commun. Phys.},
year={2021},
month={Mar},
day={25},
volume={4},
number={1},
pages={62},
issn={2399-3650},
doi={10.1038/s42005-021-00572-w},
url={https://doi.org/10.1038/s42005-021-00572-w}
}

@article{PRXQuantum.3.040330,
  title = {Optimal Cold Atom Thermometry Using Adaptive {B}ayesian Strategies},
  author = {Glatthard, Jonas and Rubio, Jes\'us and Sawant, Rahul and Hewitt, Thomas and Barontini, Giovanni and Correa, Luis A.},
  journal = {PRX Quantum},
  volume = {3},
  issue = {4},
  pages = {040330},
  numpages = {15},
  year = {2022},
  month = {Dec},
  publisher = {American Physical Society},
  doi = {10.1103/PRXQuantum.3.040330},
  url = {https://link.aps.org/doi/10.1103/PRXQuantum.3.040330}
}

@article{PhysRevA.97.063619,
  title = {Few-fermion thermometry},
  author = {P\l{}odzie\ifmmode \acute{n}\else \'{n}\fi{}, Marcin and Demkowicz-Dobrza\ifmmode \acute{n}\else \'{n}\fi{}ski, Rafa\l{} and Sowi\ifmmode \acute{n}\else \'{n}\fi{}ski, Tomasz},
  journal = {Phys. Rev. A},
  volume = {97},
  issue = {6},
  pages = {063619},
  numpages = {5},
  year = {2018},
  month = {Jun},
  publisher = {American Physical Society},
  doi = {10.1103/PhysRevA.97.063619},
  url = {https://link.aps.org/doi/10.1103/PhysRevA.97.063619}
}

@article{PhysRevLett.104.133601,
  title = {Quantum Noise Limited and Entanglement-Assisted Magnetometry},
  author = {Wasilewski, W. and Jensen, K. and Krauter, H. and Renema, J. J. and Balabas, M. V. and Polzik, E. S.},
  journal = {Phys. Rev. Lett.},
  volume = {104},
  issue = {13},
  pages = {133601},
  numpages = {4},
  year = {2010},
  month = {Mar},
  publisher = {American Physical Society},
  doi = {10.1103/PhysRevLett.104.133601},
  url = {https://link.aps.org/doi/10.1103/PhysRevLett.104.133601}
}

@article{PhysRevLett.120.260503,
  title = {Universal Quantum Magnetometry with Spin States at Equilibrium},
  author = {Troiani, Filippo and Paris, Matteo G. A.},
  journal = {Phys. Rev. Lett.},
  volume = {120},
  issue = {26},
  pages = {260503},
  numpages = {5},
  year = {2018},
  month = {Jun},
  publisher = {American Physical Society},
  doi = {10.1103/PhysRevLett.120.260503},
  url = {https://link.aps.org/doi/10.1103/PhysRevLett.120.260503}
}

@article{PhysRevLett.131.133602,
  title = {Quantum-Enhanced Magnetometry at Optimal Number Density},
  author = {Troullinou, Charikleia and Lucivero, Vito Giovanni and Mitchell, Morgan W.},
  journal = {Phys. Rev. Lett.},
  volume = {131},
  issue = {13},
  pages = {133602},
  numpages = {5},
  year = {2023},
  month = {Sep},
  publisher = {American Physical Society},
  doi = {10.1103/PhysRevLett.131.133602},
  url = {https://link.aps.org/doi/10.1103/PhysRevLett.131.133602}
}

@article{PhysRevLett.129.120503,
  title = {Sequential Measurements for Quantum-Enhanced Magnetometry in Spin Chain Probes},
  author = {Montenegro, Victor and Jones, Gareth Si\^on and Bose, Sougato and Bayat, Abolfazl},
  journal = {Phys. Rev. Lett.},
  volume = {129},
  issue = {12},
  pages = {120503},
  numpages = {7},
  year = {2022},
  month = {Sep},
  publisher = {American Physical Society},
  doi = {10.1103/PhysRevLett.129.120503},
  url = {https://link.aps.org/doi/10.1103/PhysRevLett.129.120503}
}

@article{topological_thermometry,
  title = {Topological quantum thermometry},
  author = {Srivastava, Anubhav Kumar and Bhattacharya, Utso and Lewenstein, Maciej and P\l{}odzie\ifmmode \acute{n}\else \'{n}\fi{}, Marcin},
  journal = {Phys. Rev. A},
  volume = {111},
  issue = {5},
  pages = {052216},
  numpages = {9},
  year = {2025},
  month = {May},
  publisher = {American Physical Society},
  doi = {10.1103/PhysRevA.111.052216},
  url = {https://link.aps.org/doi/10.1103/PhysRevA.111.052216}
}

@book{landau2013statistical,
  title={Statistical Physics: Volume 5},
  author={Landau, Lev Davidovich and Lifshitz, Evgenii Mikhailovich},
  volume={5},
  year={2013},
  publisher={Elsevier},
doi = {https://doi.org/10.1016/C2009-0-24487-4}
}

@book{Fischer_Hertz_1991, 
place={Cambridge}, 
series={Cambridge Studies in Magnetism}, 
title={Spin Glasses}, 
publisher={Cambridge University Press}, 
author={Fischer, K. H. and Hertz, J. A.}, 
year={1991}, 
collection={Cambridge Studies in Magnetism},
doi ={https://doi.org/10.1017/CBO9780511628771}
}

@article{wolf,
author = {Wolfgang Lechner  and Philipp Hauke  and Peter Zoller },
title = {A quantum annealing architecture with all-to-all connectivity from local interactions},
journal = {Science Advances},
volume = {1},
number = {9},
pages = {e1500838},
year = {2015},
doi = {10.1126/sciadv.1500838},
URL = {https://www.science.org/doi/abs/10.1126/sciadv.1500838}
}

@article{simon,
author = {Simon E. Nigg  and Niels Lörch  and Rakesh P. Tiwari },
title = {Robust quantum optimizer with full connectivity},
journal = {Science Advances},
volume = {3},
number = {4},
pages = {e1602273},
year = {2017},
doi = {10.1126/sciadv.1602273},
URL = {https://www.science.org/doi/abs/10.1126/sciadv.1602273}
}

@Article{helstrom1969quantum,
author={Helstrom, Carl W.},
title={Quantum detection and estimation theory},
journal={J. Stat. Phys.},
year={1969},
month={Jun},
volume={1},
number={2},
pages={231-252},
issn={1572-9613},
doi={10.1007/BF01007479},
url={https://doi.org/10.1007/BF01007479}
}

@article{PhysRevE.110.024132,
  title = {Non-Markovian enhancement of nonequilibrium quantum thermometry},
  author = {Aiache, Y. and Seida, C. and El Anouz, K. and El Allati, A.},
  journal = {Phys. Rev. E},
  volume = {110},
  issue = {2},
  pages = {024132},
  numpages = {9},
  year = {2024},
  month = {Aug},
  publisher = {American Physical Society},
  doi = {10.1103/PhysRevE.110.024132},
  url = {https://link.aps.org/doi/10.1103/PhysRevE.110.024132}
}

@article{Sekatski2022optimal,
  doi = {10.22331/q-2022-12-07-869},
  url = {https://doi.org/10.22331/q-2022-12-07-869},
  title = {Optimal nonequilibrium thermometry in {M}arkovian environments},
  author = {Sekatski, Pavel and Perarnau-Llobet, Mart{\'{i}}},
  journal = {{Quantum}},
  issn = {2521-327X},
  publisher = {{Verein zur F{\"{o}}rderung des Open Access Publizierens in den Quantenwissenschaften}},
  volume = {6},
  pages = {869},
  month = dec,
  year = {2022}
}

@article{PhysRevA.108.062421,
  title = {Invasiveness of nonequilibrium pure-dephasing quantum thermometry},
  author = {Albarelli, Francesco and Paris, Matteo G. A. and Vacchini, Bassano and Smirne, Andrea},
  journal = {Phys. Rev. A},
  volume = {108},
  issue = {6},
  pages = {062421},
  numpages = {15},
  year = {2023},
  month = {Dec},
  publisher = {American Physical Society},
  doi = {10.1103/PhysRevA.108.062421},
  url = {https://link.aps.org/doi/10.1103/PhysRevA.108.062421}
}

@Article{Razavian2019,
author={Razavian, Sholeh
and Benedetti, Claudia
and Bina, Matteo
and Akbari-Kourbolagh, Yahya
and Paris, Matteo G. A.},
title={Quantum thermometry by single-qubit dephasing},
journal={The European Physical Journal Plus},
year={2019},
month={Jun},
day={25},
volume={134},
number={6},
pages={284},
issn={2190-5444},
doi={10.1140/epjp/i2019-12708-9},
url={https://doi.org/10.1140/epjp/i2019-12708-9}
}

@article{paris2009quantum,
author = {Paris, Matteo G. A.},
title = {Quantum estimation for quantum technology},
journal = { Intl. J. Quant. Inf.},
volume = {07},
number = {supp01},
pages = {125-137},
year = {2009},
doi = {10.1142/S0219749909004839},
URL = {https://doi.org/10.1142/S0219749909004839},
}

@article{toth2014quantum,
author = {G{\'{e}}za T{\'{o}}th and Iagoba Apellaniz},
	title = {Quantum metrology from a quantum information science perspective},
	year = 2014,
	month = {oct},
	publisher = {{IOP} Publishing},
	volume = {47},
	number = {42},
	pages = {424006},
	journal = {J. Phys. A: Math. Theor.},
        doi = {10.1088/1751-8113/47/42/424006},
	url = {https://doi.org/10.1088/1751-8113/47/42/424006}
}

@Article{giovannetti2011advances,
author={Giovannetti, Vittorio and Lloyd, Seth and Maccone, Lorenzo},
title={Advances in quantum metrology},
journal={Nat. Photonics},
year={2011},
month={Apr},
day={01},
volume={5},
number={4},
pages={222-229},
issn={1749-4893},
doi={10.1038/nphoton.2011.35},
url={https://doi.org/10.1038/nphoton.2011.35}
}

@article{Mehboudi_2019,
doi = {10.1088/1751-8121/ab2828},
url = {https://dx.doi.org/10.1088/1751-8121/ab2828},
year = {2019},
month = {jul},
publisher = {IOP Publishing},
volume = {52},
number = {30},
pages = {303001},
author = {Mohammad Mehboudi and Anna Sanpera and Luis A Correa},
title = {Thermometry in the quantum regime: Recent theoretical progress},
journal = {J. Phys. A: Math. Theor.}
}

@article{campbell2018precision,
	doi = {10.1088/2058-9565/aaa641},
	url = {https://doi.org/10.1088/2058-9565/aaa641},
	year = 2018,
	month = {feb},
	publisher = {{IOP} Publishing},
	volume = {3},
	number = {2},
	pages = {025002},
	author = {Steve Campbell and Marco G Genoni and Sebastian Deffner},
	title = {Precision thermometry and the quantum speed limit},
	journal = {Quantum Sci. Technol.},
}

@article{PhysRevResearch.2.033498,
  title = {Nonequilibrium readiness and precision of Gaussian quantum thermometers},
  author = {Mancino, Luca and Genoni, Marco G. and Barbieri, Marco and Paternostro, Mauro},
  journal = {Phys. Rev. Res.},
  volume = {2},
  issue = {3},
  pages = {033498},
  numpages = {8},
  year = {2020},
  month = {Sep},
  publisher = {American Physical Society},
  doi = {10.1103/PhysRevResearch.2.033498},
  url = {https://link.aps.org/doi/10.1103/PhysRevResearch.2.033498}
}

@article{Paris_2016,
doi = {10.1088/1751-8113/49/3/03LT02},
url = {https://dx.doi.org/10.1088/1751-8113/49/3/03LT02},
year = {2015},
month = {dec},
publisher = {IOP Publishing},
volume = {49},
number = {3},
pages = {03LT02},
author = {Paris, Matteo G A},
title = {Achieving the Landau bound to precision of quantum thermometry in systems with vanishing gap},
journal = {J. Phys. A: Math. Theor. }
}

@article{Ravell,
doi = {10.1088/1367-2630/ad1d75},
url = {https://dx.doi.org/10.1088/1367-2630/ad1d75},
year = {2024},
month = {jan},
publisher = {IOP Publishing},
volume = {26},
number = {1},
pages = {013046},
author = {Ravell Rodríguez, Ricard and Mehboudi, Mohammad and Horodecki, Michał and Perarnau-Llobet, Martí},
title = {Strongly coupled fermionic probe for nonequilibrium thermometry},
journal = {New J. Phys.}
}

@article{PhysRevLett.123.180602,
  title = {Collisional Quantum Thermometry},
  author = {Seah, Stella and Nimmrichter, Stefan and Grimmer, Daniel and Santos, Jader P. and Scarani, Valerio and Landi, Gabriel T.},
  journal = {Phys. Rev. Lett.},
  volume = {123},
  issue = {18},
  pages = {180602},
  numpages = {6},
  year = {2019},
  month = {Oct},
  publisher = {American Physical Society},
  doi = {10.1103/PhysRevLett.123.180602},
  url = {https://link.aps.org/doi/10.1103/PhysRevLett.123.180602}
}

@article{bina2018continuous,
  title={Continuous-variable quantum probes for structured environments},
  author={Bina, Matteo and Grasselli, Federico and Paris, Matteo GA},
  journal={Physical Review A},
  volume={97},
  number={1},
  pages={012125},
  year={2018},
  publisher={APS}
}

@article{PhysRevA.99.062114,
  title = {Dynamical role of quantum signatures in quantum thermometry},
  author = {Feyles, Michele M. and Mancino, Luca and Sbroscia, Marco and Gianani, Ilaria and Barbieri, Marco},
  journal = {Phys. Rev. A},
  volume = {99},
  issue = {6},
  pages = {062114},
  numpages = {7},
  year = {2019},
  month = {Jun},
  publisher = {American Physical Society},
  doi = {10.1103/PhysRevA.99.062114},
  url = {https://link.aps.org/doi/10.1103/PhysRevA.99.062114}
}

@article{PhysRevLett.96.010401,
    title = {Quantum Metrology},
    author = {Giovannetti, Vittorio and Lloyd, Seth and Maccone, Lorenzo},
    journal = {Phys. Rev. Lett.},
     volume = {96},
    issue = {1},
    pages = {010401},
    numpages = {4},
    year = {2006},
    month = {Jan},
    publisher = {American Physical Society},
    doi = {10.1103/PhysRevLett.96.010401},
    url = {https://link.aps.org/doi/10.1103/PhysRevLett.96.010401}
}

@article{PhysRevLett.125.080402,
	title = {In Situ  Thermometry of a Cold {F}ermi Gas via Dephasing Impurities},
	author = {Mitchison, Mark T. and Fogarty, Thom\'as and Guarnieri, Giacomo and Campbell, Steve and Busch, Thomas and Goold, John},
	journal = {Phys. Rev. Lett.},
	volume = {125},
	issue = {8},
	pages = {080402},
	numpages = {7},
	year = {2020},
	month = {Aug},
	publisher = {American Physical Society},
	doi = {10.1103/PhysRevLett.125.080402},
	url = {https://link.aps.org/doi/10.1103/PhysRevLett.125.080402}
}

@article{PhysRevA.82.011611,
	title = {Quantum limits of thermometry},
	author = {Stace, Thomas M.},
	journal = {Phys. Rev. A},
	volume = {82},
	issue = {1},
	pages = {011611},
	numpages = {4},
	year = {2010},
	month = {Jul},
	publisher = {American Physical Society},
	doi = {10.1103/PhysRevA.82.011611},
	url = {https://link.aps.org/doi/10.1103/PhysRevA.82.011611}
}

@article{PhysRevA.98.050101,
	title = {Bridging thermodynamics and metrology in nonequilibrium quantum thermometry},
	author = {Cavina, Vasco and Mancino, Luca and De Pasquale, Antonella and Gianani, Ilaria and Sbroscia, Marco and Booth, Robert I. and Roccia, Emanuele and Raimondi, Roberto and Giovannetti, Vittorio and Barbieri, Marco},
	journal = {Phys. Rev. A},
	volume = {98},
	issue = {5},
	pages = {050101},
	numpages = {5},
	year = {2018},
	month = {Nov},
	publisher = {American Physical Society},
	doi = {10.1103/PhysRevA.98.050101},
	url = {https://link.aps.org/doi/10.1103/PhysRevA.98.050101}
}

@article{PhysRevLett.114.220405,
	title = {Individual Quantum Probes for Optimal Thermometry},
	author = {Correa, Luis A. and Mehboudi, Mohammad and Adesso, Gerardo and Sanpera, Anna},
	journal = {Phys. Rev. Lett.},
	volume = {114},
	issue = {22},
	pages = {220405},
	numpages = {5},
	year = {2015},
	month = {Jun},
	publisher = {American Physical Society},
	doi = {10.1103/PhysRevLett.114.220405},
	url = {https://link.aps.org/doi/10.1103/PhysRevLett.114.220405}
}

@article{PhysRevLett.72.3439,
	title = {Statistical distance and the geometry of quantum states},
	author = {Braunstein, Samuel L. and Caves, Carlton M.},
	journal = {Phys. Rev. Lett.},
	volume = {72},
	issue = {22},
	pages = {3439--3443},
	numpages = {0},
	year = {1994},
	month = {May},
	publisher = {American Physical Society},
	doi = {10.1103/PhysRevLett.72.3439},
	url = {https://link.aps.org/doi/10.1103/PhysRevLett.72.3439}
}
\end{document}